\documentclass[structabstract]{aa}
\usepackage{natbib}
\usepackage{url}
\usepackage{amsmath}
\usepackage{graphicx}
\usepackage{aas_macros}

\newcommand{\abs}[1]{\left\lvert #1 \right\rvert}
\newcommand{\be}{\begin{equation}}
\newcommand{\ee}{\end{equation}}
\newcommand{\ba}{\begin{align}}
\newcommand{\ea}{\end{align}}
\newcommand{\bea}{\begin{eqnarray}}
\newcommand{\eea}{\end{eqnarray}}
\newcommand{\rd}{\rm{d}}

\begin{document}
\title{Measuring the Integrated Sachs Wolfe Effect}
\author{F.-X. Dup\'e \inst{1} \thanks{francois-xavier.dupe@cea.fr} \and A. Rassat \inst{1} \thanks{anais.rassat@cea.fr} \and J.-L. Starck \inst{1}
  \and M. J. Fadili\inst{2}}
\institute{$^1$ Laboratoire AIM, UMR CEA-CNRS-Paris 7, Irfu, SAp/SEDI, Service d'Astrophysique, CEA Saclay, F-91191 GIF-SUR-YVETTE CEDEX, France. \\
  $^2$ GREYC UMR CNRS 6072, Universit\'e de Caen Basse-Normandie/ENSICAEN, 6 bdv Mar\'echal Juin, 14050 Caen, France.}
\date{} 


\abstract
{ One of the main challenges of modern cosmology is to understand the nature of the mysterious dark energy which causes
  the cosmic acceleration.  The Integrated Sachs-Wolfe (ISW) effect is sensitive to dark energy and if detected in
    a universe where modified gravity and curvature are excluded, presents an independent signature of dark energy.
  The ISW effect occurs on large scales, where cosmic variance is high and where there are large amounts of missing data
  in the CMB and large scale structure maps due to Galactic confusion. Moreover, existing methods in the literature
  often make strong assumptions about the statistics of the underlying fields or estimators.  Together these effects can
  severely limit signal extraction.}  { We want to define an optimal statistical method for detecting the ISW effect,
  which can handle large areas of missing data and minimise the number of underlying assumptions made about the data and
  estimators.}  {We first review current detections (and non-detections) of the ISW effect, comparing statistical
  subtleties between existing methods, and identifying several limitations. We propose a novel method to detect and
  measure the ISW signal.  This method assumes only that the primordial CMB field is Gaussian.  It is based on a sparse
  inpainting method to reconstruct missing data and uses a bootstrap technique to avoid assumptions about the statistics
  of the estimator.  It is a complete method, which uses three complementary statistical methods.} {We apply our
    method to Euclid-like simulations and show we can expect a $\sim7\sigma$ model-independent detection of the ISW
    signal with WMAP7-like data, even when considering missing data. Other tests return $\sim4.7\sigma$ detection levels
    for a Euclid-like survey. We find detections levels are independent from whether the galaxy field is normally or
    lognormally distributed.  We apply our method to the 2 Micron All Sky Survey (2MASS) and WMAP7 CMB data and find
    detections in the $1.1-2.0\sigma$ range, as expected from our simulations. As a by-product, we have also
  reconstructed the full-sky temperature ISW field due to 2MASS data.}  {We have presented a novel technique, based on
  sparse inpainting and bootstrapping, which accurately detects and reconstructs the ISW effect.}  

\maketitle

\section{Introduction}



The recent abundance of cosmological data in the last few decades \citep[for an example of the most recent results
see][]{Komatsu:2009, Percival:2007b,Schrabback:2010} has provided compelling evidence towards a standard concordance
cosmology, in which the Universe is composed of approximately 4\% baryons, 26\% `dark' matter and 70\% `dark' energy.

One of the main challenges of modern cosmology is to understand the nature of the mysterious dark energy which drives the observed cosmic acceleration \citep{DETF,WGFC} .

The Integrated Sachs-Wolfe (ISW) \citep{Sachs:1967er} effect is a secondary anisotropy of the Cosmic Microwave Background (CMB), which arises because of the variation with time of the cosmic
gravitational potential between local observers and the surface of last scattering.  The potential can be traced by Large Scale Structure (LSS) surveys \citep{Crittenden:1995ak}, and the ISW effect is
therefore a probe which links the high redshift CMB with the low redshift matter distribution and can be detected by cross-correlating the two.

As a cosmological probe, the ISW effect has less statistical power than weak lensing or galaxy clustering \citep[see for e.g., ][]{Euclidsb}, but it is directly sensitive to dark energy, curvature or
modified gravity \citep{Kamionkowski:1994s,Kinkhabwala:1999k,Carroll:2004de,Song:2007hs}, such that in universes where modified gravity and curvature are excluded, detection of the ISW signal provides
a direct signature of dark energy. In more general universes, the ISW effect can be used to trace alternative models of gravity.

The CMB WMAP survey is already optimal for detecting the ISW signal (see Sections \ref{sec:theory} and \ref{sec:method}), and significance is not expected to increase with the arrival of Planck, unless the
effect of the foreground Galactic mask can be reduced.  The amplitude of the measured ISW signal should however depend strongly on the details of the local tracer of mass.  Survey optimisations
\citep{Douspis:2008} show that an ideal ISW survey requires the same configuration as surveys which are optimised for weak lensing or galaxy clustering - meaning that an optimal measure of the ISW
signal will essentially come `for free' with future planned weak lensing and galaxy clustering surveys \citep[see for e.g. the Euclid survey ][]{Euclidsb}.  In the best scenario, a 4$\sigma$ detection is expected
\citep{Douspis:2008}, and it has been shown that combined with weak lensing, galaxy correlation and other probes such as clusters, the ISW can be useful to break parameter degeneracies
\citep{Euclidsb}, making it a promising probe.

Initial attempts to detect the ISW effect with COBE as the CMB tracer were fruitless \citep{Boughn:2002}, but since the arrival of WMAP data, tens of positive detections have been made, with the
highest significance reported for analyses using a tomographic combination of surveys (see Sections \ref{sec:theory} and \ref{sec:method} for a detailed review of detections).  However, several
studies using the same tracer of LSS appear to have contradicting conclusions, some analyses do not find correlation where others do, and as statistical methods to analyse the data evolve, the
significance of the ISW signal is sometimes reduced \citep[see for e.g., ][]{Afshordi:2003xu,Rassat:2007KRL,Francis:2010iswdetection}.

In Section \ref{sec:theory}, we describe the cause of the ISW effect and review current detections.  In Section \ref{sec:method}, we describe the methodology for detection and measuring the ISW signal,
and review a large proportion of reported detections in the literature, as well as their advantages and disadvantages. Having identified the main issues with current methods, we propose a new and
complete method in Section \ref{sec:saclaymethod}, which capitalises on the fact that different statistical methods are complementary and uses sparse inpainting to solve the issue of missing data and a bootstrapping technique to measure the estimator's probability distribution function (PDF).  In Section \ref{sec:validation}, we validate our new method using simulations for 2MASS and Euclid-like surveys. In
Section \ref{sec:data}, we apply our new method to WMAP 7 and the 2MASS survey. In Section \ref{sec:discussion}, we present our conclusions.

\section{The Integrated Sachs-Wolfe Effect}\label{sec:theory}
\subsection{Origin of the Integrated Sachs-Wolfe Effect} \label{sec:theory:origin} 

General relativity predicts that the wavelength of electromagnetic radiation is sensitive to gravitational potentials, an effect which is called \emph{gravitational redshift}.  Photons travelling from
the surface of last scattering will necessarily travel through the gravitational potential of Large Scale Structure (LSS) on their way to the observer; these will be blue-shifted as they enter the
potential well and red-shifted as they exit the potential.  These shifts will accumulate along the line of sight of the observer.  The total shift in wavelength will translate into a change in the
measured temperature-temperature anisotropy of the CMB, and can be calculated by: \begin{equation} \left(\frac{\Delta T}{T}\right)_{ISW}=-2\int_{\eta_L}^{\eta_0}\Phi'\left((\eta_0-\eta)\hat{\bf
      n},\eta \right)\rd\eta, \label{sec:theory:eq:isw} \end{equation} where $T$ is the temperature of the CMB, $\eta$ is the conformal time, defined by $\rd \eta = \frac{\rd t}{a(t)}$ and $\eta_0$
and $\eta_L$ represent the conformal times today and at the surface of last scattering respectively.  The unit vector $\hat{\bf n}$ is along the line of sight and the gravitational potential
$\Phi({\bf x}, \eta)$ depends on position and time.  The integral depends on the rate of change of the potential $\Phi'=\rd \Phi / \rd\eta$.

In a universe with no dark energy or curvature, the cosmic (linear) gravitational potential does not vary with time, so that such a blue- and red-shift will always cancel out, because $\Phi' = 0$ and
there will be no net effect on the wavelength of the photon.

However, in the presence of dark energy or curvature \citep{Sachs:1967er,Kamionkowski:1994s,Kinkhabwala:1999k}, the right hand side of Equation \ref{sec:theory:eq:isw} will be non-null as the cosmic
potential will change with time \citep[see for e.g., ][]{Dodelson:2003}, resulting in a secondary anisotropy in the CMB temperature field.

\begin{table*}[htbp]         
   \centering
   
      \begin{tabular}{@{} lllllc @{}} 
\hline
\hline
Author   & CMB& LSS Tracer &Wavelength&Method&Claimed\\
&&&&& Detection\\
                \hline         
\cite{Boughn:2002}&COBE &XRB&Xray&D2&No\\
  \cite{Giannantonio:2008}&W3&&&D2&2.7$\sigma$\\
\hline
         \cite{Boughn:2003yz,Boughn:2004zm} &W1&XRB/NVSS& Xray/Radio&D2 &`tentative' (2-3 $\sigma$)\\
              \hline 
      \raisebox{-1.8ex}{\cite{Fosalba:2003ge}}&\raisebox{-1.8ex}{W1}&\raisebox{-1.8ex}{SDSS DR1}&&\raisebox{-1.8ex}{D2}&\raisebox{-1.5ex}{2$\sigma$ (low z)}\\
      &&&&&\raisebox{1ex}{3.6$\sigma$ (high z)}\\
      \cite{Cabre:2006qm}&W3&SDSS DR4&Optical &D2&$>2\sigma$\\
                \cite{Giannantonio:2008}&W3& SDSS DR6&&D2&2.2$\sigma$\\
                    \cite{Shanks:isw}&W5&SDSS DR5&&D2&`marginal'\\
                                            \cite{Lopez:2010nocross} &W5&SDSS DR7&&D2&`No detection'\\
\hline
                                                                                             \cite{Giannantonio:2006al}&W3&SDSS Quasars &Optical&D2&2$\sigma$\\
                                                                                             \cite{Giannantonio:2008}&W3& SDSS Quasars&&D2&2.5$\sigma$\\
                                                                                             \cite{Xia2009}&W5&SDSS Quasars&&D2&2.7$\sigma$\\
                                                                                             \hline
                            \cite{Scranton:2003in} & W1 & & & D2 & $>2\sigma$ \\
                             \cite{Padmanabhan:2004fy} & W1&&&D1 &  2.5$\sigma$\\	
                                                      \cite{Granett:2009}&W3&SDSS LRG&Optical&D1&$2 \sigma$\\
\cite{Giannantonio:2008}&W3& &&D2&2.2$\sigma$\\
&&&&&\\
                                      \cite{Shanks:isw}&W5&SDSS LRG, 2SLAQ&&D2&`marginal'\\
                   \cite{Shanks:isw}&W5&AAOmega LRG&&D2&Null\\

                                                      \hline
\cite{Fosalba:2004ge}&W1&APM&Optical &D2&2.5$\sigma$\\
\hline
\cite{Afshordi:2003xu}&W1&&&D1& 2.5 $\sigma$\\
\cite{Rassat:2007KRL}&W3&2MASS&NIR&D1&2$\sigma$\\	
\cite{Giannantonio:2008}&W3& &&D2&0.5$\sigma$\\
\cite{Francis:2010iswdetection}&W3&&&D1&`weak'\\
 \hline
    \cite{Boughn:2002}&COBE & &&D2&No\\

\cite{Nolta:2003uy} &W1& &  &D2&2.2$\sigma$\\			
                        \cite{Pietrobon:2006}&W3& NVSS& Radio&D3&$>4\sigma$\\	
                        \cite{Vielva2006} & W3 & & & D3 & $3.3\sigma$ \\
                                                \cite{McEwen:2006md} &W3&& &D3&$>2.5\sigma$\\	
                       \cite{Raccanelli2008} & W3& &  & D2 & $2.7 \sigma$ \\
                        \cite{McEwen:2008}&W3& & & D3& $\sim 4\sigma$\\
                          \cite{Giannantonio:2008}&W3& &&D2&3.3$\sigma$\\
                        \cite{Hernandez:2009}&W3&&&D1&$<2\sigma$\\
                                                                    \cite{Shanks:isw}&W5&&&D2&`marginal' ($\sim 2\sigma$)\\

\hline

\cite{Corasaniti:2005}&W1&&&D2&$>2\sigma$\\
\cite{Gaztanaga:2004sk}&W1&&&D2& 2$\sigma$\\
\cite{Ho:2008}&W3&\raisebox{0.5ex}{Combination}&\raisebox{0.5ex}{Combination}&D1&$3.7 \sigma$\\
\cite{Giannantonio:2008}&W3& &&D2&4.5$\sigma$\\
\hline
      
     \end{tabular}
    
     \begin{center}

\caption{Meta-analysis of ISW detections to date and their reported statistical significance.  The `Method' describes the space in which the power spectrum analysis is done (configuration, spherical harmonic, etc \ldots), not the method for measuring the significance level of the detection (this is described in Section \ref{sec:method}). {\bf D1} corresponds to spherical harmonic space, {\bf D2} to configuration space, {\bf D3} to wavelet space.  The highest detections are made in wavelet space.  Regarding the survey used, the highest detections are made using NVSS (though weak and marginal detections using NVSS are also reported) or using combinations of LSS surveys as the matter tracer.}
      \label{sec:theory:tab:detections}
   \end{center}
\end{table*}
\subsection{Detection of the ISW signal}\label{sec:theory:detections}

The ISW effect leads to a linear scale secondary anisotropy in the temperature field of the CMB, and will thus affect the CMB temperature power spectrum at large scales.  Due to the primordial anisotropies and cosmic variance on large scales, the ISW signal is difficult to detect directly in the temperature map of the CMB, but \cite{Crittenden:1995ak} showed it could be detected through cross-correlation of the CMB with a local tracer of mass.

The first attempt to detect the ISW effect \citep{Boughn:2002} involved correlating the Cosmic Microwave Background
explorer data \citep[][COBE]{COBE:1990} with XRB \citep{Boldt1987} and NVSS data \citep{Condon:1998}.  This analysis did
not find a significant correlation between the local tracers of mass and the CMB.  Since the release of data from the
Wilkinson Microwave Anisotropy Probe \cite[][WMAP]{Spergel:2003cb} over $20$ studies (see Table
\ref{sec:theory:tab:detections}) have investigated cross-correlations between the different years of WMAP data and local
tracers selected using various wavelengths: X-ray \cite[][XRB survey]{Boldt1987}; optical \citep[][SDSS galaxies]{SDSS:optical2006,SDSS:2008},
\citep[][SDSS QSOs]{SDSS:qso2001}, \citep[][SDSS LRGs]{SDSS:lrg2004}, \cite[][APM]{APM:1990}; near infrared \citep[][2MASS]{Jarrett:2000me}; radio
\citep[][NVSS]{Condon:1998}.

The full sky WMAP data have sufficient resolution on large scales that the measure of the ISW signal is cosmic variance limited.   The best LSS probe of the ISW effect should include maximum sky coverage and full redshift coverage of the dark energy dominated era \citep{Douspis:2008}.  No such survey exists yet, so there is room for improvement on the ISW detection as larger and larger LSS surveys arise.  For this reason, when we review the current ISW detections, we classify them according to their tracer of LSS, and not the CMB map used.

The measure of the ISW signal can be done in various statistical spaces; we classify detections in Table \ref{sec:theory:tab:detections} into three measurement `domains': D1 corresponds to spherical harmonic space; D2 to configuration space and D3 to wavelet space. (In Section \ref{sec:method}, we review the different methods for quantifying the statistical significance of each measurement).

There are only two analyses which use COBE as CMB data \citep[with XRB and NVSS data,][]{Boughn:2002}, and both report null detections, which can reasonably be due to the low angular resolution of COBE even at large scales.  The rest are done correlating WMAP data from years 1, 3 and 5 (respectively `W1', `W3' and `W5' in table \ref{sec:theory:tab:detections}).
  
Most ISW detections reported in Table \ref{sec:theory:tab:detections} are relatively `weak' ($<3\sigma$) and this is expected from theory for a concordance cosmology.   Higher detections are reported for the NVSS survey \citep{Pietrobon:2006,McEwen:2008,Giannantonio:2008}, though weak and marginal detections using NVSS data are also reported \citep{Hernandez:2009,Shanks:isw}.  High detections are often made using a wavelet analysis \citep{Pietrobon:2006,McEwen:2008}, though a similar study by the same authors using the same data but a different analysis method finds a weaker signal \citep{McEwen:2006md}.  The highest detection is reported using a tomographic combination of all surveys \citep[XRB, SDSS galaxies, SDSS QSOs, 2MASS and NVSS, ][]{Giannantonio:2008}, as expected given the larger redshift coverage of the analysis.

Several analyses have been revisited to seek confirmation of previous detections.  In some cases, results are very similar (\cite{Padmanabhan:2004fy,Granett:2009,Giannantonio:2008}, for SDSS LRGs; \cite{Giannantonio:2006al,Giannantonio:2008} for SDSS Quasars; \cite{Afshordi:2003xu,Rassat:2007KRL}, for 2MASS), but in some cases they are controversially different (for e.g. \cite{Pietrobon:2006} and \cite{Shanks:isw}, for NVSS or  \cite{Afshordi:2003xu} and \cite{Giannantonio:2008}, for 2MASS).  

We also notice that as certain surveys are revisited, there is a trend for the statistical significance to be reduced: for e.g., detections from 2MASS decrease from a $2.5\sigma$ detection \citep{Afshordi:2003xu}, to $2\sigma$ \citep{Rassat:2007KRL}, to $0.5\sigma$ \citep{Giannantonio:2008} to `weak' \citep{Francis:2010iswdetection}.  Detections using SDSS LRGs decrease from $2.5\sigma$ \citep{Padmanabhan:2004fy}, to $2-2.2\sigma$ \citep{Granett:2009,Giannantonio:2008}, to `marginal' \citep{Shanks:isw}.  Furthermore, there tends to be a `sociological bias' in the interpretation of the confidence on the signal detection. The first detections interpret a $2-3 \sigma$ detection as `tentative'
\citep{Boughn:2003yz,Boughn:2004zm}, while further studies with similar detection level report `independent evidence of dark energy' \citep{Afshordi:2003xu,Gaztanaga:2004sk}.

\section{Methodology for Detecting the ISW Effect}\label{sec:method}
In this paper, we are interested in qualifying the differences between different statistical methods which exist in the literature, and comparing them with a new method we present in Section \ref{sec:saclaymethod}.  By statistical method, we mean the method which is used to quantify the significance of a signal, not the space in which the signal is measured. Therefore, and without loss of generality, the review presented in Section \ref{sec:method:review} summarises methods using spherical harmonics.  We compare the pros and cons of each method in Section \ref{sec:method:vs}.  We begin by describing how the ISW signal can be measured in spherical harmonics in Section \ref{sec:method:isw}

\subsection{ISW Signal in Spherical Harmonics}\label{sec:method:isw}

In general, any field can be decomposed by a series of functions which form an orthonormal set, as do the spherical
harmonic functions $Y_{\ell m}(\theta,\phi)$.  Therefore a projected galaxy overdensity ($\delta_g$) or temperature
anisotropy ($\delta_T$) field $\delta_{X}(\theta,\phi)$, where $X = g, T$, can be decomposed into: \begin{equation}
  \delta_X(\theta,\phi)=\sum_{\ell,m}a^X_{\ell m}Y_{\ell m}(\theta,\phi),\end{equation} where $a^X_{\ell m}$ are the
spherical harmonic coefficients of the field.  The 2-point galaxy-temperature cross-correlation function can then be
written:
\begin{equation} C_{gT}(\ell)=\frac{1}{(2\ell+1)}\sum_m \mathcal{R}e\left[a^g_{\ell m} (a^{T}_{\ell m})^*\right],\end{equation} where taking the real part of the product ensures that $C_{gT} (\ell)= C_{Tg}(\ell)$.

The theory for the angular cross-correlation function is given by: 
\begin{equation}C_{gT}(\ell) = 4 \pi b_g\int \rd k \frac{\Delta^2(k)}{k}  W_g(k)W_T(k),\label{eq:cgt}\end{equation}
where 
\begin{equation}W_g(k) =\int \rd r \Theta(r) j_\ell(kr) D(z),\end{equation}
\begin{equation}W_T(k) = -\frac{3\Omega_{m,0} H_0^2}{k^2c^3}\int_0^{z_L} \rd r j_\ell(kr)H(z)D(z)(f-1),\end{equation}
\begin{equation}\Delta^2(k) = \frac{4\pi}{(2\pi)^3}k^3P(k),\end{equation}
\begin{equation}\Theta(r) = \frac{r^2n(r)}{\int \rd r r^2 n(r)}.\end{equation}
In these equations, $r$ represents the co-moving distance, $z_L$ the redshift at the last scattering surface, $k$
the Fourier mode wavenumber and quantities which depend on the redshift $z$ have an intrinsic dependence on $r$: $H(z) =
H(z(r))$.  The function $f$ is the linear growth factor given by $f=\frac{\rd \ln D(z)}{\rd \ln a(z)}$, where
$D(z)$ is the linear growth which measures the growth of structure. The cross-correlation function depends on the
survey selection function given by $n(r)$, in units of galaxies per unit volume. The quantities $\Omega_{m,0}$ and
$H_0$ are the values of the matter density and the Hubble parameter at $z=0$. Units are chosen so that the quantity
$C(\ell)$ is unitless.

In the case where both the temperature and the galaxy fields behave as Gaussian random fields, then the covariance on the ISW signal can be calculated by: 
\begin{equation}\left<\left|C_{gT}\right|^2\right> =\frac{1}{f_{sky}(2\ell+1)}\left[C^2_{gT}+\left(C_{gg}+\mathcal{N}_g\right)\left(C_{TT}+\mathcal{N}_T\right)\right],\label{eq:covar}\end{equation} where $C_{TT}$ is the temperature-temperature power spectrum, $\mathcal{N}_{g}$ and $\mathcal{N}_T$ are the noise of the galaxy and temperature fields respectively.  The galaxy auto-correlation function can be calculated theoretically in linear theory by: 
\begin{equation}C_{gg}(\ell) = 4 \pi b^2_g\int \rd k \frac{\Delta^2(k)}{k}  \left[W_g(k)\right]^2, \label{eq:cgg}\end{equation}

There are many difficulties in measuring the ISW effect, the first being the intrinsic weakness of the signal.  To add to this, an unknown galaxy bias scales linearly with the ISW cross-correlation signal (see Equation
\ref{eq:cgt}), which is therefore strongly degenerate with cosmology.  Galactic foregrounds in both the CMB and the LSS maps also mask crucial large scale data and can introduce spurious correlations.
Any method claiming to detect the ISW effect should be as thorough as possible in accounting for missing data, and where possible the reported detection level should be independent of an assumed
cosmology.

\subsection{Review on Current Tools for ISW Detection}\label{sec:method:review}

In the literature there are two quantities which can be used to measure and detect the ISW signal, which we review in this section.  Without loss of generality, we present these methods in spherical
harmonic space.  The first method measures the observed cross-correlation spectra (`Spectra' method: see section \ref{sec:method:powerspectra}), whilst the second directly compares temperature fields
(`Fields' method: see section \ref{sec:method:field}). These two approaches differ by the quantity they measure to infer a detection.  For each method (fields vs. spectra), it is possible to use
different statistical methods to infer detection which we describe below.  We review each existing method below and summarise the pros and cons of both of these classes as well as the statistical
models in Table \ref{tab:prosandcons}.

\begin{table*}
  \centering
  \begin{tabular}{|l|c|c|c|}
 \hline
 \hline
{\bf Measured} & {\bf Advantage} & {\bf Disadvantage}\\
{\bf Quantity}&&\\
  \hline

 &Methods exist for calculation with missing data. & Assumes $C(\ell)$'s or estimator are Gaussian.\\
Spectra&Can introduce tomography easily.& Most methods requires \\
 & & estimation of the covariance matrix. \\
 \hline

 Fields & No assumption about galaxy/matter density field &Missing data is an ill-posed problem.\\
 \hline 
 \hline
{\bf   Statistical}& {\bf Advantage}& {\bf Disadvantage}\\
{\bf Method}&&\\
\hline
 Simple correlation&Independent of cosmology.&Measure of significance\\
 (Spectra / Fields) &&assumes estimator is Gaussian.\\
 \hline
 Amplitude estimation& &Detection depends on cosmology/model. \\
(Spectra / Fields) &Validates signal and model simultaneously.& Measure of significance assumes estimator is Gaussian.\\
& &Assumes underlying theory (e.g. $\Lambda$CDM) is correct. \\
  \hline  
 $\chi^2$ (Spectra / Fields) & &Assumes $C(\ell)$'s are Gaussian (Spectra).\\
  & Validates signal and model simultaneously.&Only gives confidence of rejecting null hypothesis.\\
& &Assumes underlying theory (e.g. $\Lambda$CDM) is correct. \\
 \hline
 Model comparison&Asks a different question than other tests.&ISW signal usually too weak to be detected this way.\\
 (Spectra)&&Assumes underlying theory (e.g. $\Lambda$CDM) is correct.\\

\hline 
     \end{tabular}
  \caption{TOP: Review of advantages and disadvantages of measuring spectra vs. fields in order to infer an ISW detection.  BOTTOM: Review of statistical methods \emph{existing in the literature} and their respective advantages and disadvantages.}
  \label{tab:prosandcons}
\end{table*}

\subsubsection{Note on the confidence score} \label{sec:method:confidence} 

Before reviewing the ISW detection methods, we would like to clarify the definition of confidence scores from a
  statistical point of view. The confidence of a null hypothesis test can be interpreted as the distance from the data
  to the null hypothesis (commonly named $H_0$). For example, let $\rho$ be a variable of interest (e.g. correlation
  coefficient, amplitude). The confidence score $\sigma$ for the hypothesis test $H_0$ (i.e., $\rho = 0$) against $H_1$
  (i.e., $\rho \neq 0$) is directly computed using the formula $\tau = \rho/\sigma(\rho)$, where $\sigma(\rho)$ is the
  standard deviation. But this is only true when 1) $\rho$ is Gaussian, 2) $\sigma(\rho)$ is computed independently from the
  observation and 3) considering a symmetric test.  Then, this method does not stand for the general case and as the
  correlation coefficient considered here must be positive, an asymmetric (one-sided) test would be more
  appropriate here.

As the confidence score is directly linked to the deviation from the $H_0$ hypothesis through the $p$-value, the $\sigma$-score is always positive. Then the $p$-value $p$ of one hypothesis test is
  computed using the probability density function (PDF) of the test distribution: $p = 1 - \int_{-\infty}^{\tau}
  P(x|H_0) \mathrm{d}x = \int_{\tau}^{+\infty} P(x|H_0) \mathrm{d}x$ (for a classical one-sided test). Remark that in
  that case, if the $H_0$ is true then $\rho \approx 0$ (i.e. in the middle of the test distribution) and the $p$-value
  $p$ will be around $0.5$ which correspond to a confidence of $0.67\sigma$.  

\subsubsection{Note on the application spaces} \label{sec:method:applicsp}

All the methods that will be described in the next sections can be performed in different
domains. While some spaces may be more appropriate than others for a specific task,
difficulties may also arise because of the properties of the space. For example: for the
`spectra' method in configuration space, the two main difficulties are missing data and
the estimation of the covariance matrix \citep[see e.g.][]{Hernandez:2008}.  In this case,
the covariance matrix can be estimated using Monte Carlo methods
\citep[see][]{Cabre:2007}.  In spherical harmonic space, missing data induce mode
correlations which can be removed by using an appropriate framework for calculation of the $C(\ell)$'s
\citep[e.g.][]{Hivon:2002}. In harmonic space (when missing data is accounted for) the covariance matrix is diagonal and thus easily invertible (see Section \ref{sec:method:field}).

\subsubsection{Cross-power spectra comparison}\label{sec:method:powerspectra}
The most popular method consists in using the cross-correlation function (Equation
\ref{eq:cgt}, in spherical harmonic space) to measure the presence of the ISW signal,
however this approach has recently been challenged by \citet{Lopez:2010nocross}, because
of its high sensitivity to noise and fluctuations due to cosmic variance.

One of the subtleties of the cross-correlation function method is the evaluation of the covariance matrix
$C_{\mathrm{covar}}$ and its inverse. This matrix can be estimated using the MC1 or MC2 methods of \citet{Cabre:2007},
in which case the test is strongly dependent on the quality of the simulations. Secondly, missing data will require
extra care when estimating the power spectra, this can be tackled by using MASTER (Monte Carlo Apodized Spherical Transform Estimator) or QML (Quadratic Maximum Likelihood) methods
\citep{Hivon:2002,Efstathiou2004hybrid,Munshi2009gauss}.

The spectra measurement can be used with one of four different statistical methods. The advantages and disadvantages of each method are summarised below and in Table \ref{tab:prosandcons}. The
first aims to detect a correlation between two signals, i.e. we test if the cross-power
spectra is null or not. The second fits a (model-dependent) template to the measured cross-power spectra.
The other two methods aim to validate a cosmological model as well as confirm the presence of
a signal: the $\chi^2$ test and the model comparison.  We describe them below:

\begin{itemize}
\item{\bf Simple correlation detection:} The simplest and the most widely used method for detecting a
  cross-correlation between two fields $X$ and $Y$ (here supposing that $Y$ is correlated with $X$) \citep[see
  e.g.][]{Boughn:2002,Afshordi:2003xu,Pietrobon:2006,Shanks:isw} is to measure the correlation coefficient $\rho(X,Y)$, defined as:
  \begin{equation}
    \label{eq:9}
    \begin{split}
      \rho(X,Y) = \mathrm{Cor}(X,Y) / \mathrm{Cor}(X,X), \\
      {\rm where,~}
      \mathrm{Cor}(X,Y) = \frac{1}{N_p} \sum_p\mathcal{R}e\left[ X^*(p) Y(p)\right]~,
    \end{split}
  \end{equation}
  with $p$ a position or scale parameter and $N_p$ the number of considered positions or scales. There is a
  correlation between the two fields if $\rho(X,Y)$ is not null. The correlation coefficient
  is linked to the cross-power spectra in harmonic space:
  \begin{equation}
    \label{eq:10}
    \rho(g,T) = \left( \sum_{\ell} (2\ell + 1) C_{gT}(\ell) \right) / \left( \sum_{\ell} (2\ell + 1) C_{gg}(\ell) \right)~.
  \end{equation}
  Thus the nullity of the coefficient implies the nullity of the cross-power spectra.  A z-score can be performed in order to test this nullity,
  \begin{equation}
    \label{eq:19}
    K_0 = \rho / \sigma_{\rho}~,
  \end{equation}
  where the standard error of the correlation value, $\sigma_{\rho}$ can be estimated using Monte Carlo simulations
  under a given cosmology.  In the literature, most applications of this method assume that $\boldsymbol{K_0}$ follows a
  Gaussian distribution under the null hypothesis (i.e. no correlation), for example~\cite{Vielva2006} and
  \cite{Giannantonio:2008}, which is not necessarily true. 

  The distribution of the $K_0$ test can be inferred if we assume that both fields $X$ and $Y$, of the cross-correlation
  are Gaussian. In this case, the correlation coefficient distribution is the normally distributed but follows a normal
  product distribution, which is far from Gaussian. In the case where $Y$ is a constant field, the correlation
  coefficient follows a normal distribution and the distribution of the hypothesis test $K_0$ will depend on how
    the variance of the estimator $\sigma_{\rho}$ is computed. If this last value is derived from the observation $X$,
    then $K_0$ follows a Student's t-distribution which converges to a Gaussian distribution only when $\rho$ is high
    (by the central limit theorem). Otherwise, if the variance $\sigma_{\rho}$ is estimated independently from the data
    (through Monte-Carlo, for example) or known for a given cosmology, $K_0$ can be assumed to follow a Gaussian
    distribution. This means that $K_0$ is generally not Gaussian even if the correlation coefficient is Gaussian (see
    section~\ref{sec:saclaymethod}).

  This method does not include knowledge of the underlying ISW signal, nor of the galaxy field, though the error bars
  can be estimated from Monte Carlo simulations which include cosmological information.

\item{\bf Amplitude estimation (or template matching):}

  The principle of the amplitude estimation is to measure whether an observed signal corresponds to the signal predicted
  by a given cosmological model. 
  
  The estimator and its variance are given by  \citep[e.g.][]{Ho:2008,Giannantonio:2008}:
\begin{equation}
  \label{eq:3}
  \hat{\lambda} = \frac{C_{gT}^{\mathrm{Th}*} {C}_{\mathrm{covar}}^{-1} C_{gT}^{\mathrm{Obs}}}{C_{gT}^{\mathrm{Th}*} C_{\mathrm{covar}}^{-1} C_{gT}^{\mathrm{Th}}}~, \quad
  \sigma_{\hat{\lambda}} = \frac{1}{\sqrt{C_{gT}^{\mathrm{Th}*} C_{\mathrm{covar}}^{-1} C_{gT}^{\mathrm{Th}}}}~,
\end{equation} 
where $C^{\mathrm{Th}}_{gT}$ is the theoretical cross-power spectrum,
$C^{\mathrm{Obs}}_{gT}$ the estimated (observed) power spectrum and $C_{\rm covar}$ the
covariance matrix calculated by Equation \ref{eq:covar}.  A z-score,
\begin{equation}
  \label{eq:15}
  K_1 = \hat{\lambda} / \sigma_{\hat{\lambda}}~,
\end{equation}
is usually applied to test if the amplitude is null or not.

\item{\bf Goodness of fit,  $\chi^2$ test:}

The goodness of fit or $\chi^2$ test is given by \citep[e.g.][]{Afshordi:2003xu,Rassat:2007KRL}:
\begin{equation}
  \label{eq:4}
  K_2 = (C_{gT}^{\mathrm{Th}} - C_{gT}^{\mathrm{Obs}} )^* {C}_{\mathrm{covar}}^{-1} (C_{gT}^{\mathrm{Th}} - C_{gT}^{\mathrm{Obs}} )~,
\end{equation}
where $K_2$ follows a $\chi^2$ distribution with number of degrees of freedom (\emph{d.o.f}) depending on the input
data. This tests
the correspondance of the data with a given cosmological model, but does not infer if the tested model is in fact the
best. Equation \ref{eq:4} also assumes that the $C(\ell)'s$ are Gaussian variables. The value of $\chi^2$ gives an idea on the probability of rejecting the model, but cannot be compared directly with
the $K_2$ value for the null hypothesis without careful statistics (see next method on model comparison).  
\item{\bf Model comparison:}\label{spectra:model}

  The model comparison method is based on the generalised likelihood ratio test and asks the question: \emph{`Do
    the data prefer a given fiducial cosmological model over the null hypothesis ?'}.  This question is important because
  it could be possible to use the previous $\chi^2$ test to detect an ISW signal - yet the data could still also be
  compatible with a null hypothesis \citep[see for e.g.,][]{Afshordi:2003xu,Rassat:2007KRL,Francis:2010iswdetection}. In this case it
  is important to perform a model comparison to find which model is preferred by the data. Two hypotheses
  are built:
\begin{itemize}
\item $H_0$: \emph{``there is no ISW signal''}, i.e. the cross-power spectra is null;
\item $H_1$: \emph{``there is an ISW signal compatible with a fiducial cosmology''}, i.e. the cross-power spectra is close to an expected one.
\end{itemize}

One should then estimate:\begin{equation}
  \label{eq:5}
  \begin{split}
    K_3 &= \Delta \chi^2= (C_{gT}^{\mathrm{Obs}} )^* {C}_{\mathrm{covar}}^{-1} (C_{gT}^{\mathrm{Obs}} ) - \\
    &(C_{gT}^{\mathrm{Th}} - C_{gT}^{\mathrm{Obs}} )^* {C}_{\mathrm{covar}}^{-1} (C_{gT}^{\mathrm{Th}} - C_{gT}^{\mathrm{Obs}} )~,
  \end{split}
\end{equation} where $K_3$ converges asymptotically to a $\chi^2$ distribution. If the value is higher than a threshold (chosen for a
required confidence level), the $H_0$ hypothesis is rejected.  However, such method is difficult to use directly because
of the small sample bias, $K_3$ is not likely to follow a $\chi^2$ statistic.  In the case of the ISW signal, the signal
for a `standard' fiducial cosmology (e.g., WMAP 7 cosmology) is so weak that it usually returns a lack of detection for
current surveys - this may not be the case for future or tomographic surveys. Notice that this model comparison method
can be seen as an improved version of the goodness of fit.

\end{itemize}

\subsubsection{Field to field comparison}\label{sec:method:field}

Instead of comparing the spectra, one can work directly with the temperature field to measure the presence of the
ISW signal. The observable in this case is now the ISW temperature field ($\delta_{\mathrm{ISW}}$), rather than the cross-correlation power spectra $C_{gT}(\ell)$.
The observed CMB temperature anisotropies $\delta_{\rm OBS}$ can be described as: \begin{equation}
  \label{eq:7}
  \delta_{\mathrm{OBS}} = \delta_{T} + \lambda \delta_{\mathrm{ISW}}+\delta_{\mathrm{other}}+\mathcal{N}~,
\end{equation}
where $\delta_{\mathrm{ISW}}$ is the ISW field and $\lambda$ its amplitude (normally near 1), $\delta_T$ the primordial
CMB temperature field, $\delta_{\rm other}$ represents fluctuations due to secondary anisotropies other than the ISW
effect and $\mathcal{N}$ represents noise.  In the context of the ISW effect, which occurs only on large (linear) scales
where noise and other secondary anisotropies are negligible, we have:
\begin{equation}
  \delta_{\mathrm{OBS}} \simeq \delta_{T} + \lambda \delta_{\mathrm{ISW}}~.
\end{equation}

The main difference between the fields and spectra approach is that the fields method requires an estimation of
the ISW temperature field ($\delta_{\rm ISW}$). There are several methods to calculate $\delta_{\rm ISW}$ from a given
matter overdensity map. The most accurate way to reconstruct the ISW signal is to use information from the full 3-dimensional matter distribution, which in theory requires overlapping galaxy and weak lensing maps on large scales, in order to measure the galaxy bias.  This may be possible in the future with surveys like Euclid \citep{Euclidsb}. Assuming a simple bias relation, the matter field can also be estimated directly from galaxy surveys \citep[see][who did this for small patches on the sky]{Granett:2009}.  In the case where only the general redshift distribution of the galaxy survey is known, the ISW field $\delta_{\rm ISW}$ can be approximated directly from the galaxy and temperature maps
using \citep[see][]{Boughn:1998,Cabre:2007,Giannantonio:2008}:
\begin{equation}
  \label{eq:8}
  a^{\rm ISW}_{\ell m} = \frac{C_{gT}(\ell)}{C_{gg}(\ell)} g_{\ell m}~,
\end{equation}

where $g_{\ell m}$ are the spherical harmonic coefficients of the galaxy map, and $a^{\rm ISW}_{\ell m}$ the
coefficients of the ISW temperature anisotropy map. 

Another approach is to reconstruct the ISW map using
Equation~\ref{sec:theory:eq:isw} where $\Phi'$ is estimated using the Poisson Equation \citep{Francis:2010iswanomalies}.

In general, it is assumed that on non-linear scales, the ISW is called the Rees-Sciama effect and will produce a negatively correlated signal due to non-linear growing modes of the matter distribution \citep{Schaefer:2010,Cai:2010}.  However, some non-linear modes could also be decaying for example due to major mergers or tidal stripping, or due to alternative cosmologies as in \cite{Afshordi:2011}. In this case the signal could be positively correlated even on non-linear scales (from equation \ref{sec:theory:eq:isw}). The total ISW signal (i.e., the signal which is positively correlated) would not necessarily be Gaussian in this case.

In our approach, we do not model possible contributions from non-linear (growing or decaying) modes, but we allow quasi-linear modes in the data or simulations to produce a positively correlated ISW signal as an approximation; we do this since it is in practice very difficult to separate linear and quasi-linear modes.

As for the spectra approach, there are several statistical methods available to qualify detection: 
\begin{itemize}

\item{\bf Simple correlation detection:}

  The simple correlation detection method described for the spectra comparison, can in fact also be considered as a
  field comparison. This is the only method which directly overlaps between both approaches.

\item {\bf Amplitude estimation (or template matching)}:

Using the Gaussian framework, given an ISW field, the amplitude $\lambda$ can be
estimated with the corresponding maximum likelihood estimator \citep{Hernandez:2008,Frommert:2008,Granett:2009}: \begin{equation}
  \label{eq:1}
  \hat{\lambda} = \frac{\delta_{\mathrm{ISW}}^* C_{TT}^{-1} \delta_{\mathrm{OBS}}}{\delta_{\mathrm{ISW}}^* C_{TT}^{-1} \delta_{\mathrm{ISW}}}, \quad
  \sigma_{\hat{\lambda}} = \frac{1}{\sqrt{\delta_{\mathrm{ISW}}^* C_{TT}^{-1} \delta_{\mathrm{ISW}}}}~.
\end{equation}
A signal is present if $\hat{\lambda}$ is non null and a z-score,
\begin{equation}
  \label{eq:16}
  K_4 = \hat{\lambda} / \sigma_{\hat{\lambda}}~,
\end{equation}
directly yields the confidence level in terms of $\sigma$.  Equation \ref{eq:1} implicitly
assumes that the primordial CMB field $\delta_T$ is a Gaussian random field (we discuss this further in section \ref{sec:saclaymethod}). 

\item{\bf Goodness of fit, $\chi^2$ test:}

The $\chi^2$ goodness of fit with $H_1$ (see {\bf Model comparison} in \ref{spectra:model}) yields:
\begin{equation}
  \label{eq:2}
  K_5 = (\delta_{\mathrm{ISW}} - \delta_{\mathrm{OBS}})^* C_{TT}^{-1}  (\delta_{\mathrm{ISW}} - \delta_{\mathrm{OBS}})~,
\end{equation}
where $K_5$ is a $\chi^2$ variable with number of \emph{d.o.f} depending on the
input data.  In this case, the test only returns the confidence of rejecting the null hypothesis $H_0$.  As in the $\chi^2$ test for the spectra, precaution must be taken when comparing $\chi^2$ values for different models, by using an appropriate model comparison technique. We introduce this in section \ref{sec:saclaymethod}.

\end{itemize}

\subsection{Pros and cons of each method}\label{sec:method:vs}

We have identified two main classes of methods: either using power spectra or fields to measure the ISW signal.  For
each approach one can choose amongst several statistical tools to measure the significance of a correlation or validate
simultaneously a correlation and a model.  The advantages and disadvantages of both approaches are summarised below and
in the top part of Table \ref{tab:prosandcons}.

One of the main advantages of using the field approach is that it assumes only that the primordial CMB field comes from
a Gaussian random process, which is largely believed to be true.  In the other approach, the spectra are assumed to be
Gaussian, which is not the case. Several studies \citep{Cole:2001,Suto:2001,Wild:2005} have also shown that the matter
overdensity exhibits a lognormal behavior on large scale. The bias introduced has been shown to be small
\citep{Bernardeau:2002,Hamimeche:2008}, however this approach is still theoretically ill-motivated. The main advantage
of the spectra method is the relative ease when calculating the spectra from incomplete data sets, as tools are
available for calculating the spectra \citep[see e.g.][]{Efstathiou2004hybrid}. In the field approach, managing missing
data is an ill-posed problem.

We remind that a problem is defined as a well-posed problem \citep{Hadamard1902} if 1) a solution exists, 2) the
solution is unique and 3) the solution depends continuously on the data (in some reasonable topology). Otherwise, the
problem is defined as a ill-posed problem. With missing data, the second point cannot be verified. Reconstruction of the
data also requires inversion of an operation (e.g., the mask), and is therefore an \emph{ill-posed inverse problem}.
Notice that most inverse problems are ill-posed (e.g. deconvolution).

\section{A rigorous method for detecting the ISW effect}\label{sec:saclaymethod}

Having identified in the previous section the numerous methods used in the literature to detect and measure the ISW
signal, as well as their relative advantages and disadvantages, we propose here a complete and rigorous method for
detecting and quantifying the signal significance. We describe this method in detail below and it is summarised in
Figure \ref{fig:saclaymethod}.
 
\subsection{Motivation}
\label{sec:motivation}

If we consider the pros and cons of each detection method shown in Table \ref{tab:prosandcons} and in Section
\ref{sec:method:vs}, we remark first that any method based on the comparison of spectra makes the demanding assumption
that the $C(\ell)'s$ be Gaussian, whereas methods based on field comparison require only the primordial CMB field to be
Gaussian.  Instead the fields method assumes only that the primordial CMB is Gaussian, so we recommend this method be used for an ISW analysis. 

\begin{figure}[htbp]
   \centering
   \includegraphics[width=8cm]{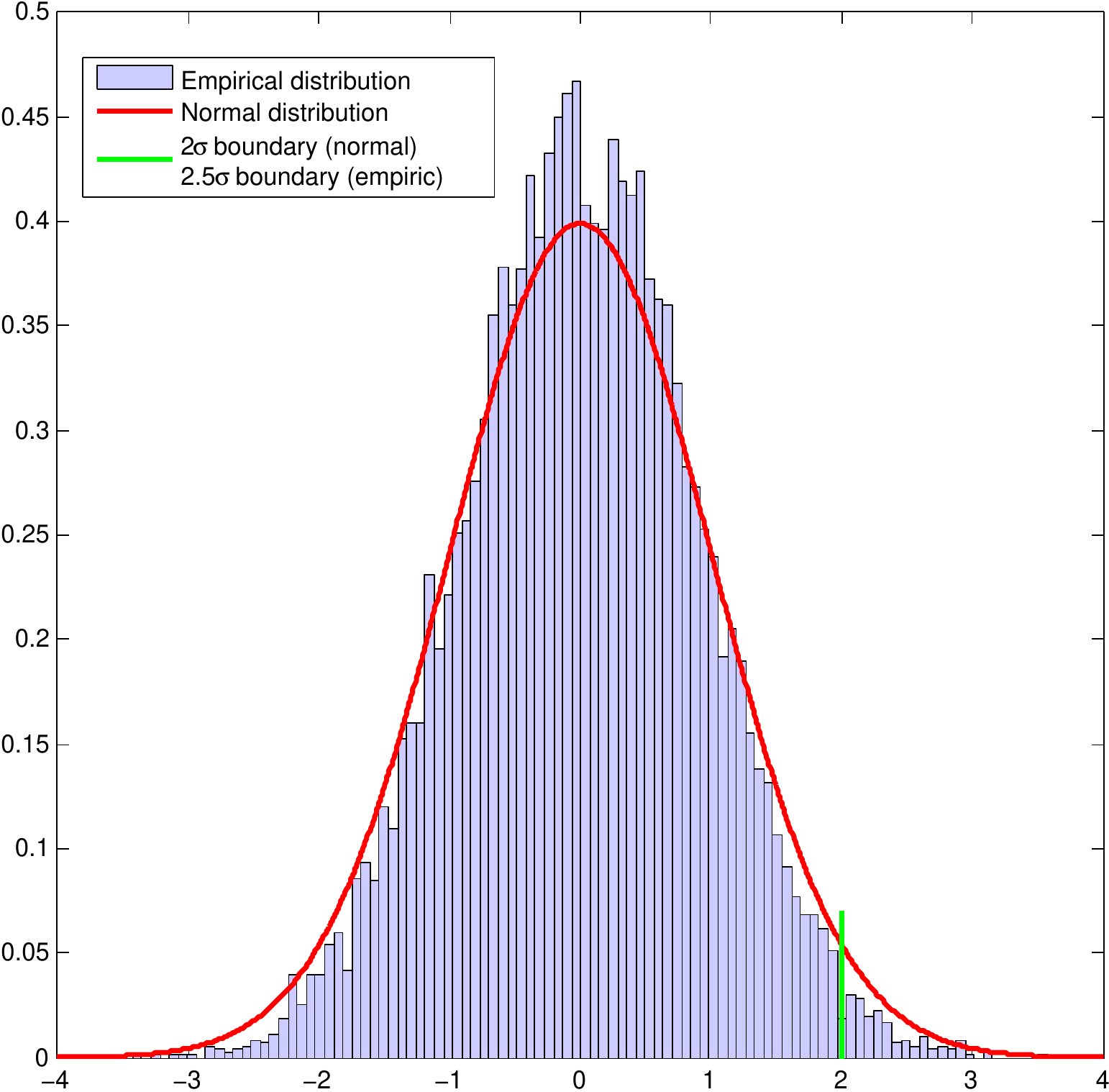}
   \caption{Comparison of Gaussian PDF for estimator $\lambda/\sigma_\lambda$ with its true estimated distribution. A
     $2\sigma$ significance using the Gaussian PDF corresponds in fact to a $2.5\sigma$ detection with the true
     distribution. (Calculation for 2MASS survey, see section \ref{sec:data}).}
   \label{fig:edf}
\end{figure}

Another key issue is the estimation of the signal significance.  We find that most approaches in the literature assume
that the probability distribution function (PDF) of the \emph{estimator} is Gaussian.  In Figure \ref{fig:edf} we
evaluate the estimator's PDF (under the null hypothesis) for the ISW signal due the 2MASS survey (see Section
\ref{sec:data}) using a Monte-Carlo method (purple bars) and compare it with a Gaussian PDF (red solid line).  The
distributions' tails differ which leads to a bias in the confidence level for positive data (i.e. data with an ISW
signal). For the 2MASS survey, a 2$\sigma$ detection with a Gaussian assumption (vertical solid/green line) corresponds
in fact to a $2.5\sigma$ detection using the true underlying PDF.  In this case, the signal amplitude is underestimated
with the Gaussian assumption. Such behavior has also been studied for marginal detections by \citet{Bassett2010}.  To
avoid this bias, we recommend that the PDF be estimated, and not assumed Gaussian.

Finally, there is not one `ideal' statistical method.  Different methods have both advantages and disadvantages and a
combination of different methods can prove complementary.

\subsection{The Saclay method}
\label{sec:saclay-method}

\begin{figure*}[htbp]
   \centering
   \includegraphics[width=15cm]{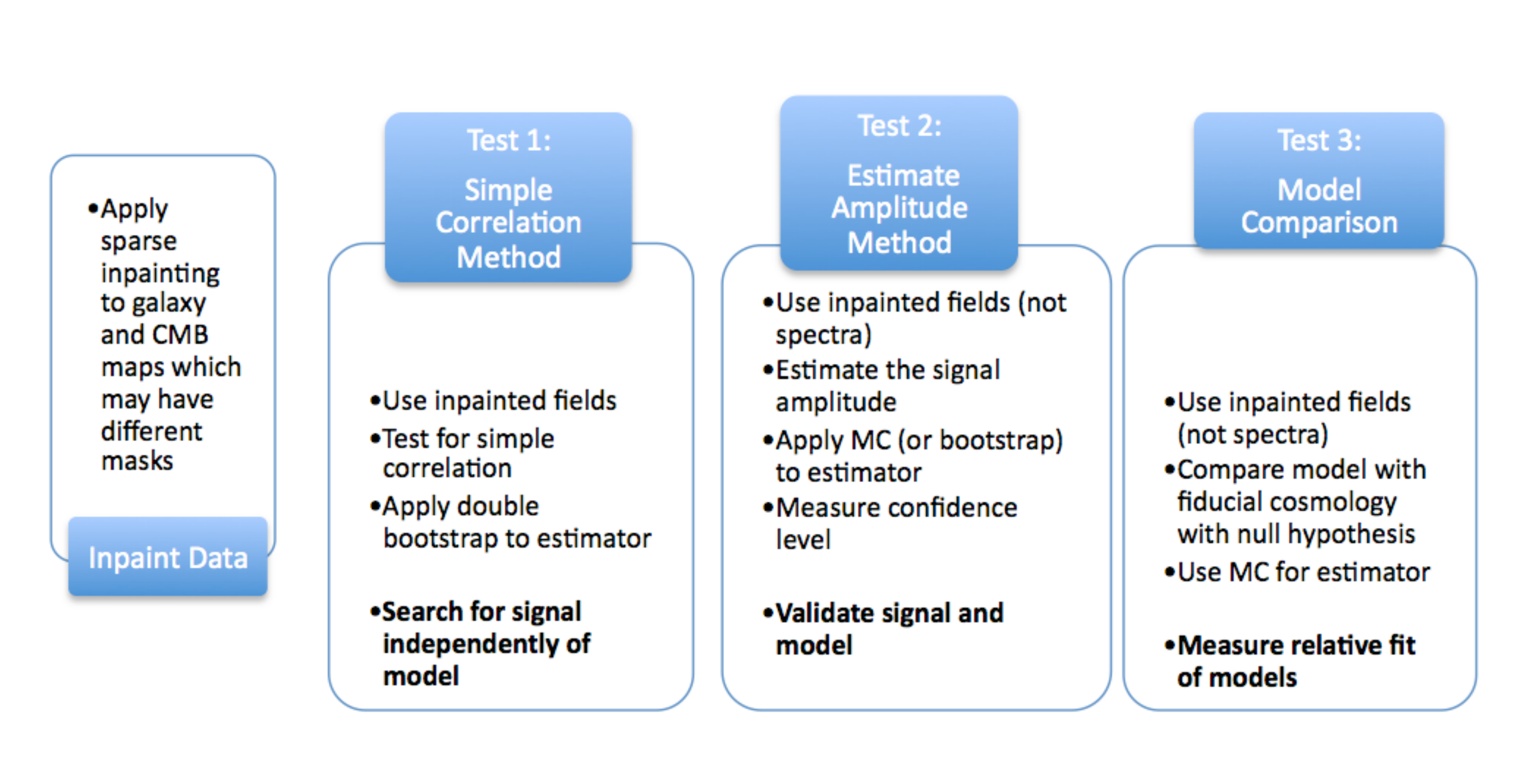} 
   \caption{Description of the steps involved in our method for detecting the ISW effect.  Tests 1 to 3 are complementary and ask different statistical questions.}
   \label{fig:saclaymethod}
\end{figure*}

We present here a new ISW detection method which uses the fields as input (and not the spectra).  As we have seen,
several statistical tests are interesting in the sense that they do not address the same questions. Therefore we believe
that a solid ISW detection method should test:
\begin{enumerate}
\item \textit{The correlation detection}: this test is independent of the cosmology. 
\item \textit{The amplitude estimation}: this will seek for a specific signal.
\item \textit{The model comparison}: this allows us to check whether the model with ISW is preferred to the model
  without ISW.
\end{enumerate}

Using the fields instead of the spectra means we must deal with the problem of missing data,  which we solve with a
sparse inpainting technique (see Appendix \ref{app:inpainting}). Such a method has already been applied with success for CMB lensing
estimations~\citep{perotto10}.

The last important issue is how we estimate the final detection level. As explained before, the z-score asymptotically follows a
Gaussian distribution and so a bootstrap or Monte-Carlo method is required to derive the correct $p$-value from the \textit{true} test distribution.

\subsubsection{Methodology}
\label{sec:methodology}

Our optimal strategy (summarised by Figure~\ref{fig:saclaymethod}) for ISW detection is the following:
\begin{enumerate}
\item Apply sparse inpainting to both the galaxy and CMB maps which may have different masks, essentially reconstructing missing data around the Galactic plane and bulge. 
\item Test for simple correlation, using a double bootstrap (one to estimate the variance and another to estimate the confidence) and using the fields as input. This returns a model-independent detection level. 
\item Reconstruct the ISW signal using expected cosmology and the inpainted galaxy density map.
\item Estimate the signal amplitude using the fields as input, and apply a bootstrap (or MC) to the estimator. This validates both the signal and the model.
\item Apply the model comparison test using the fields as input (see Section \ref{sec:fields-models-comp}), essentially testing whether the data prefers a fiducial (e.g., $\Lambda$CDM) model over the null hypothesis, and measure relative fit of models.
\end{enumerate}

As we have chosen to work with the fields, the very first step is to deal with the missing data.  This is an ill-posed
problem, which can be solved using sparse inpainting (see Appendix \ref{app:inpainting} for more details). This approach
reconstructs the entirety of the field, including along the Galactic plane and bulge. We show in Section \ref{sec:validation} that the use of sparse inpainting does not introduce a bias in the detection of the ISW effect.

We then perform a correlation detection on the reconstructed fields data using a double bootstrap (see
Appendix~\ref{app:bootstrap:correlation}).  
Experiments show that the bootstrap tends to over-estimate the
confidence interval, especially when the $p$-values are small. This is why the obtained detection must be used as a
indicator when near a significant value (for high $p$-values, the bootstrap remains accurate).

The second test evaluates the signal amplitude,
which validates both the presence of a signal and the chosen model.  Bootstrapping test can also be used here
as it has no assumption on the underlying cosmology. However, since the accuracy of bootstrap depends on the quantity of
observed elements, it may become inaccurate for low $p$-value, i.e. when there is detection. In such case, Monte-Carlo
(MC) will provide more accurate $p$-values and, for example, with $10^6$ MC simulations, we have an accuracy of about
$1/1000$. 

This second test compares the ISW signal with a fiducial model, but does not consider the possibility that
the measured signal could in fact be consistent with the `null hypothesis'. So even with a significant signal, a third
test is necessary.  This more pertinent question is addressed by using the `Model Comparison' method (defined in Section
\ref{sec:fields-models-comp}), for the first time using the fields approach.

In conclusion, our method consists of a series of complementary tests which together answer several questions. The first
test seeks the presence of a correlation between two fields, without any referring cosmology. The second model-dependent
test searches a given signal and tests its nullity. The third test asks whether the data prefers a fiducial ISW signal over the null hypothesis.

\subsubsection{`Field' model comparison}
\label{sec:fields-models-comp}
We define here the model comparison technique using the fields approach, which has until now not been used in the literature. Using a generalised
likelihood ratio approach, the quantity to measure is:
\begin{equation}
  \label{eq:6}
  \begin{split}
    K_6 &=  \delta_{\mathrm{OBS}}^* {C}_{TT}^{-1} \delta_{\mathrm{OBS}} - \\
    & (\delta_{\mathrm{ISW}} - \delta_{\mathrm{OBS}} )^* {C}_{TT}^{-1} (\delta_{\mathrm{ISW}} - \delta_{\mathrm{OBS}} )~.
  \end{split}
\end{equation}
Theoretically $K_6$ convergences asymptotically to a $\chi^2$ variable with a certain number of \emph{d.o.f}'s. As we
only have one observation we cannot assume (asymptotic) convergence. We can however use a Monte Carlo approach in order to estimate the $p$-value of the test under the $H_0$ hypothesis.

The $p$-value is defined as the probability that under $H_0$ the test value can be over a given $K_6$, i.e.  $P( t >
K_6) = \int_{K_6}^{\infty} p(x) dx$, where $p$ is the probability distribution of the test under the $H_0$
hypothesis. By simulating primordial CMB for a fiducial cosmology, these values can be easily computed. Then the $p$-value gives
us a confidence on rejecting the $H_0$ hypothesis.

Notice that the same procedure for the $p$-value estimation can be applied on $K_4$ (Equation \ref{eq:16}), even on
$K_1$ (Equation \ref{eq:15}) and $K_3$ (Equation \ref{eq:5}) for the power spectra methods. We will further refer to this $p$-value
estimation as the Monte-Carlo estimation, as we theoretically know the distribution under the null hypothesis, i.e. the
primordial CMB is supposed to come from a Gaussian random process.

\section{Validation of the Saclay Method}
\label{sec:validation}

In order to validate the Saclay method, we estimate the detection level expected using WMAP 7 data for the CMB and 2MASS and Euclid data for the galaxy data (see
section~\ref{sec:data} for a description of WMAP and 2MASS data sets). We quantify the effect of the inpainting process on CMB
maps with and without an ISW signal.  We do this by simulating 2MASS-like and Euclid-like Gaussian and lognormal galaxy distributions and WMAP7-like Gaussian CMB maps (using cosmological parameters from Table~\ref{tab:wmap7}) both
with and without an ISW signal.  We then apply our method to attempt a detection of the ISW signal.  We do this both on
full-sky maps as well as on masked data where we have reconstructed data behind the mask using the sparse inpainting
technique (the masks we use are as described in \ref{sec:data:wmap} and \ref{sec:data:2mass}).

For each simulation, we run the 3-step Saclay method. Except
for the cross-correlation method where we use 100 iterations for the 2MASS-like and 1000 for Euclid-like simulations for the $p$-value estimation and 201 for the variance
estimation (nested bootstrap), every other Monte-Carlo process was performed using 10 000 iterations
All tests were performed inside the spherical harmonics domain with ${\ell \in
  [2,100]}$ for 2MASS and $\ell=[2,350]$ for a Euclid-like survey.
  
For the Euclid-like survey, we consider a galaxy distribution as defined in \cite{Amara:2007}, with mean redshift $z_m=0.8$ and slopes $\alpha = 2, \beta=1.5$. We reconstruct the ISW effect created by the projected galaxy distribution of the Euclid survey, by considering only one large redshift bin. In the future, it could be possible to refine such a reconstructed map by considering tomographic bins, or using information from the spectroscopic survey. As sky coverage maps are not yet available for Euclid, we consider the same mask as for 2MASS and inpaint regions with missing data following Section \ref{sec:data:inpainting}. We choose do to this, rather than simply assume a value for the fraction of sky covered ($f_{\rm sky}$), so as to consider more realistic problems relating to the shape of the mask, and to test our inpainting method.

\begin{table*}[t]
  \centering
  {\bf
  \begin{tabular}{|l|l||c|c||c|c|}
  \hline
  \hline 
   &2MASS / Method & Full sky & Inpainted & Full sky & Inpainted  \\
  &&maps  (with ISW) &maps  (with ISW) & map (no ISW) & map (no ISW)\\
  \hline \hline
  \multicolumn{6}{|c|}{Gaussian matter density field} \\
  \hline \hline
  1& Simple correlation (b) &$1.12 \pm 0.98$  & $1.08 \pm 0.99$ & $0.78 \pm 0.81$ & $0.81 \pm 0.86$ \\
  2&Field amplitude  (MC) & $0.94 \pm 0.67$  & $0.91 \pm 0.65$ & $0.68 \pm 0.57$ & $0.71 \pm 0.58$ \\
  3&Field model test (MC) & $0.94 \pm 0.67$  & $0.91 \pm 0.64$ & $0.69 \pm 0.57$ & $0.72 \pm 0.58$ \\
  \hline

  \multicolumn{6}{|c|}{Lognormal matter density field} \\
  \hline \hline
  1& Simple correlation (b) & $1.37 \pm 1.51$ & $1.30 \pm 1.37$ & $0.85 \pm 1.05$ & $0.98 \pm 1.34$ \\
  2&Field amplitude  (MC)  & $0.97 \pm 0.62$ & $1.00 \pm 0.65$ & $0.69 \pm 0.54$ & $0.78 \pm 0.61$ \\
  3&Field model test (MC ) & $0.97 \pm 0.62$ & $1.00 \pm 0.65$ & $0.69 \pm 0.53$ & $0.78 \pm 0.60$ \\
  \hline

   \hline
  \end{tabular}}
\caption{Expected detection level (units of $\sigma$) of ISW signal for a 2MASS-like local tracer of mass assuming Gaussian and lognormal distribution for a fiducial cosmology (see Table \ref{tab:wmap7}). Methods $1-3$ represent the 3-step method. Angular scales included in the analysis are $\ell =[2-100]$.}
  \label{tab:simu2mass}
\end{table*}

\begin{table*}[t]
  \centering
  {\bf
  \begin{tabular}{|l|l||c|c||c|c|}
  \hline
  \hline 
   &Euclid / Method & Full sky & Inpainted & Full sky & Inpainted  \\
  &&maps  (with ISW) &maps  (with ISW) & map (no ISW) & map (no ISW)\\
  \hline \hline
  \multicolumn{6}{|c|}{Gaussian matter density field} \\
  \hline \hline
  1& Simple correlation (b) &$>7$  & $6.98\pm2.30$ & $0.87\pm0.67$ & $0.99\pm0.72$ \\
  2&Field amplitude  (MC) & $>5$  & $4.70\pm2.40$ & $0.78\pm0.49$ & $0.88\pm0.54$ \\
  3&Field model test (MC) & $>5$  & $4.77\pm2.42$ & $0.78\pm0.48$ & $0.88\pm0.54$ \\
  \hline
   \multicolumn{6}{|c|}{Lognormal matter density field} \\
  \hline \hline
  1& Simple correlation (b) & $>7$ & $6.29\pm2.60$ & $0.84\pm0.70$ & $0.77\pm0.72$ \\
  2&Field amplitude  (MC)  & $>5$ & $4.43\pm2.37$ & $0.75\pm0.54$ &$0.74\pm0.54$\\
  3&Field model test (MC ) & $>5$ &  $4.43\pm2.37$ & $0.74\pm0.53$&$0.74\pm0.54$ \\

   \hline
  \end{tabular}}
\caption{Expected detection level (units of $\sigma$) of ISW signal for a Euclid-like local tracer of mass assuming Gaussian distribution for a fiducial cosmology (see Table \ref{tab:wmap7}). Methods $1-3$ represent the 3-step method. Angular scales included in the analysis are $\ell =[2-300]$.}
  \label{tab:simuEuclid}
\end{table*}

\subsection{Expected level of detection:}
\label{sec:expect-level-detect}

The expected detection levels (in units of $\sigma$) are reported in Table \ref{tab:simu2mass} (2MASS) and Table \ref{tab:simuEuclid} (Euclid).  Methods $1-3$ correspond to the 3-step method described in Figure \ref{fig:saclaymethod}, where $(b)$ and $(MC)$  denote bootstrap and Monte Carlo evaluations of the variance and the $p$-value of the test. The $p$-values are converted as a $\sigma$ value using the following formula: \begin{equation}
  \label{eq:23}
  s = \sqrt{2}\ \mathrm{erf}^{-1}(1 - p)~,
\end{equation}
where $p$ is the $p$-value, $s$ the corresponding $\sigma$-score and $\mathrm{erf}^{-1}$ the inverse error function.

The 2MASS simulations (Table \ref{tab:simu2mass}) show that we expect the same level of significance for an ISW
  detection, whether the mass tracer follows a Gaussian or a lognormal distribution. In either case the significance is
  low, around $~1 \sigma \pm 1\sigma$. This means that for 2MASS-like survey, we have a signal to noise ratio (S/N)
  around $1\sigma$.  We see no major difference between the expected detection levels of M2 and M3.  The only difference
  is for M1 (but there is still agreement with M2 and M3 within $1\sigma$ error bars) - this may be due to the fact that
  the bootstrap technique is more efficient for Gaussian
  assumptions.

  We also apply our method to CMB simulations with no ISW signal present (2 left columns of Table \ref{tab:simu2mass}),
  and find a lower detection significance than when an ISW signal is present.  This is true even when inpainting is used
  to recover missing data, showing that the inpainting method does not introduce spurious correlations.

  In any case, all methods suggest it is difficult to detect the ISW signal with high significance using the 2MASS data
  as a local tracer of the matter distribution.

\begin{table}[htp]
  \centering
  {\bf
  \begin{tabular}{|p{0.4\linewidth}|c|c|}
    \hline
    \hline
    \multicolumn{3}{|c|}{Gaussian density} \\
    \hline
     Method & Monte-Carlo & Bootstrap \\
    \hline Cross-Correlation (Equation \ref{eq:9}) &  $0.46 \pm 0.28$ & $0.44 \pm 0.35$ \\
    \hline Amplitude estimation (Equation \ref{eq:1}) & $0.44 \pm 0.30$ & $0.47 \pm 0.33$ \\
    \hline
    \hline
    \multicolumn{3}{|c|}{Lognormal density} \\
    \hline
     Method & Monte-Carlo & Bootstrap \\
    \hline Cross-Correlation (Equation \ref{eq:9}) &  $0.39 \pm 0.27$ & $0.38 \pm 0.31$ \\
    \hline Amplitude estimation (Equation \ref{eq:1}) & $0.41 \pm 0.26$ & $0.40 \pm 0.30$ \\
    \hline
  \end{tabular}
}
  \caption{Expected $p$-values for inpainted maps of ISW signal for a 2MASS-like local tracer of mass using
    Monte-Carlo or bootstrap methods for the first two steps of the Saclay method.}
  \label{tab:pvalues}
\end{table}

In Table \ref{tab:simuEuclid}, we show that an Euclid-like survey, which is optimally designed for an ISW detection \citep[see,][]{Douspis:2008}, permits a much higher detection than with a 2MASS survey. As with 2MASS simulations, we notice that M2 and M3 return similar detection levels, which are lower than M1.  Inclusion of masked data reduced the significance, but our inpainting method does not introduce spurious correlations as inpainted maps with no ISW do not return a detection. For a Euclid-like survey with incomplete sky coverage, we can expect to show that the data prefers an ISW component over no dark energy (M3) at the $4.7 \sigma$ level, and detect a cross-correlation signal at the $\sim 7 \sigma$ level. We find no significant differences in the detection levels when the simulations are assumed lognormal or Gaussian.

We also investigate the performance of the wild bootstrap method for the confidence estimation. Table~\ref{tab:pvalues}
shows the $p$-values estimated using Monte-Carlo procedure and wild bootstrap for the first two methods of the 3-step
Saclay method. Notice that the bootstrap results are almost equivalent to Monte-Carlo ones. We find the bootstrap
method wasn't always reliable when the $p$-value becomes small, because the precision of the bootstrap depends on both the number of
bootstrap samples (as any MC-like process) and the number of observed elements. This last dependence makes the
bootstrap uncertain when the detection is almost certain, that is why we consider the bootstrapped cross-correlation as
an indicator that needs refinement when the results are very significant.

\subsection{Power of the tests}
\label{sec:power-tests}

\begin{figure*}[htbp]
   \centering
\includegraphics[width=8cm]{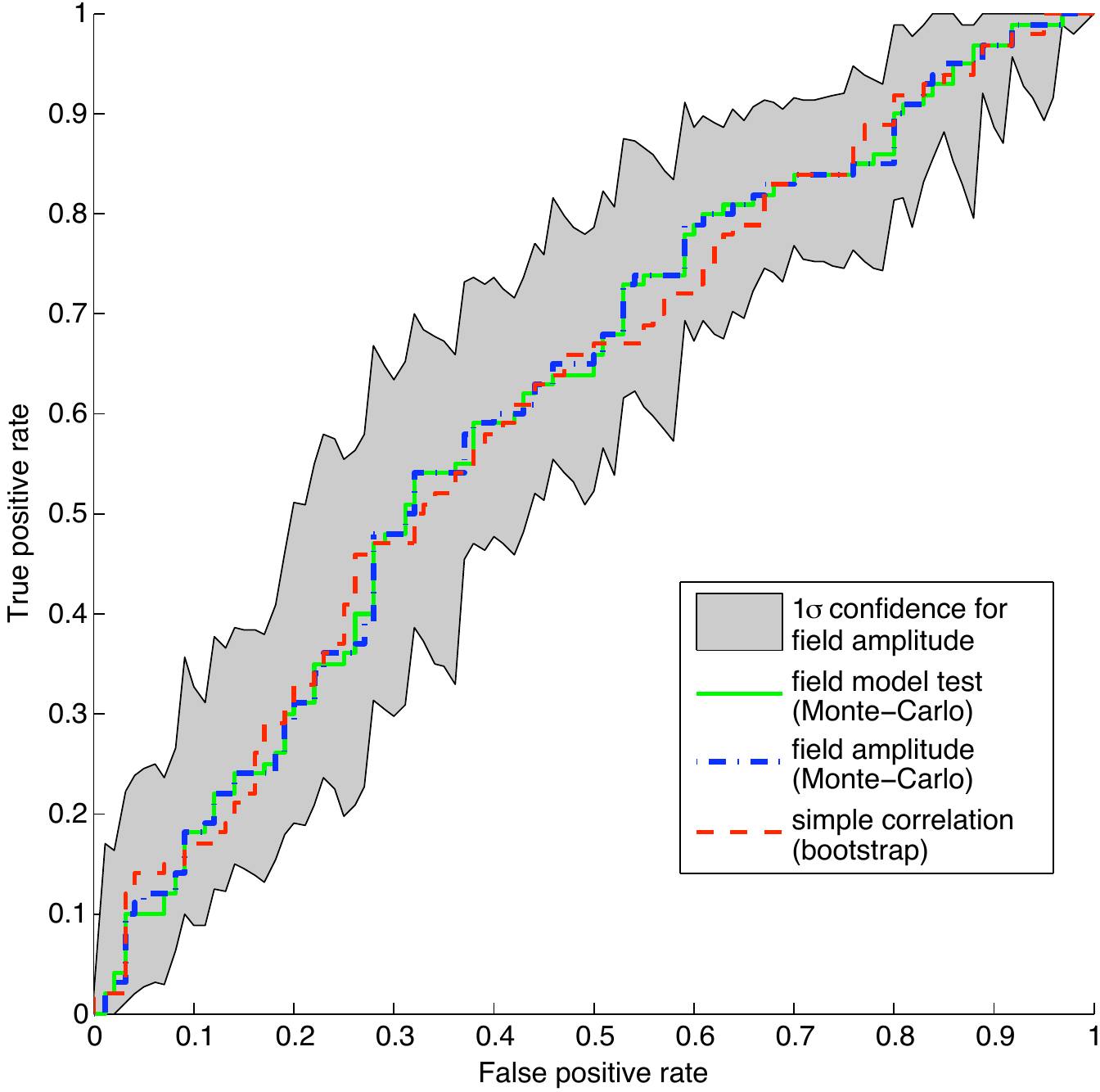}\includegraphics[width=8cm]{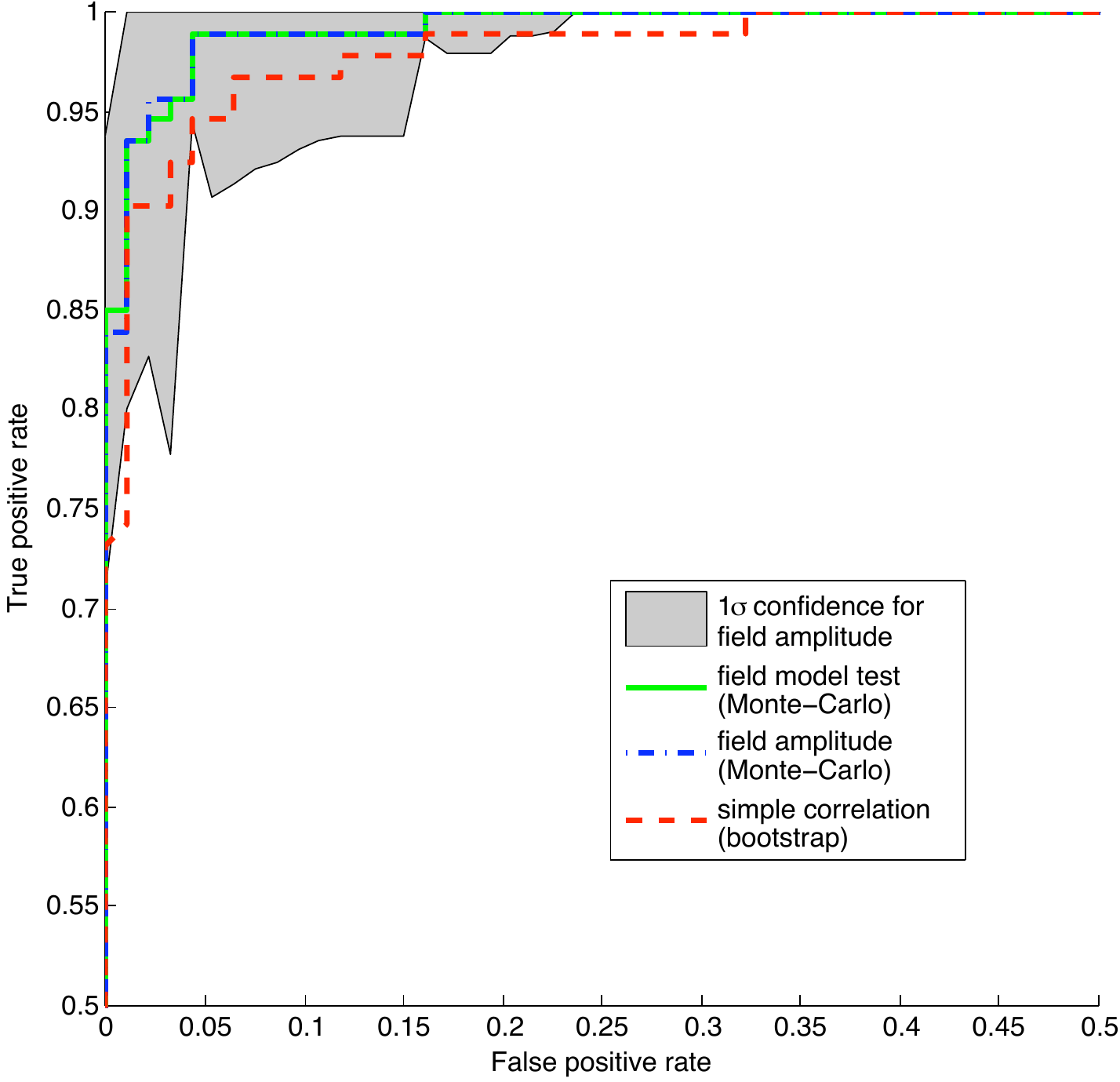} 

\caption{ROC curves for the 3-step Saclay method for the 2MASS survey (\emph{Left}) and the Euclid survey
    (\emph{Right}). Note the axes in the right-hand panel are different than in the left-hand panel. The statistical
  methods correspond to: fields model test (thin solid green, Method 3 in Table \ref{tab:simu2mass}), field amplitude
  estimation (dot-dashed blue, Method 2), simple correlation (dashed red, Method 1). For the 2MASS survey, all the
  methods are inside the $1\sigma$ error bar of the field's amplitude estimation and so are nearly equivalent.  For
    the Euclid-like survey, the statistics return much better values than for 2MASS (i.e., the ROC curves are far from
    the diagonal).  The simple correlation method will return more false positives than the other two methods, which are
    nearly identical. The ROC curves of the simple correlation test differs sometimes by more than $1\sigma$ at some
    points from the other two methods and so is expected to perform differently. }
   \label{fig:roc}
\end{figure*}

In order to investigate the different strengths of each method, we also evaluate the rate of true positives vs. the
number of false positives (i.e. false detections) - this information is summarised in Figure \ref{fig:roc} which shows
Receiver Operating Characteristic (ROC) curves.  The construction of the ROC curve requires the computation of
$p$-values for several simulated cases (i.e., simulations with and without an ISW signal), which are then sorted by
value. For each $p$-value or threshold, the corresponding false positive and true positive rates are computed.

Generally, a more sensitive method may be more permissive and so will return a higher proportion of false detections.
An ideal method will have a ROC curve above the diagonal from $(0,1)$ to $(1,1)$. Similarly, a poor detector will
produce a curve below the diagonal, which corresponds to odds worse than tossing a coin. The X-axis corresponds to
  the false positive rate, i.e. the ratio of CMB maps without ISW where ISW signal is detected at the current
  threshold. The Y-axis corresponds to the true positive rate, i.e. the ratio of CMB maps with ISW where ISW signal is
  detected at the current threshold. We recall that a point on the ROC curve corresponds to a threshold. 

Figure \ref{fig:roc} shows the ROC curves for three methods applied to a 2MASS-like survey (\emph{Left}) and a Euclid-like survey (emph{Right}): fields model test (thin solid green),
field amplitude estimation (dot-dashed blue), simple correlation (dashed red).  
For the 2MASS survey, all the methods are inside the $1\sigma$ error bar of the field's amplitude estimation and so are nearly equivalent, i.e. no method performs better than the others.  
     
For the Euclid-like survey, the statistics return much better values then for 2MASS (i.e., the ROC curves are far from the diagonal).  The simple correlation method will return more false positives than the other two methods, which are nearly identical. The ROC curves for each method differ by more than $1\sigma$ at some points and so different methods will perform differently.

\section{The ISW signal in WMAP7 due to 2MASS galaxies}
\label{sec:data}

We apply the new detection method described in Section \ref{sec:saclaymethod} to WMAP7 data \citep{Jarosik2010} and the
2MASS galaxy survey which has been extensively used as a tracer of mass for the ISW signal (see Table
\ref{sec:theory:tab:detections}).
We describe first the data in Sections \ref{sec:data:wmap} and \ref{sec:data:2mass}.  In Section \ref{sec:data:inpainting} we describe the inpainting process that we apply to both CMB and galaxy data. In Section \ref{sec:results}, we present the detection results.

\subsection{WMAP}\label{sec:data:wmap}

\begin{figure*}[htbp]
   \centering
   \includegraphics[width=7.5cm,angle=90]{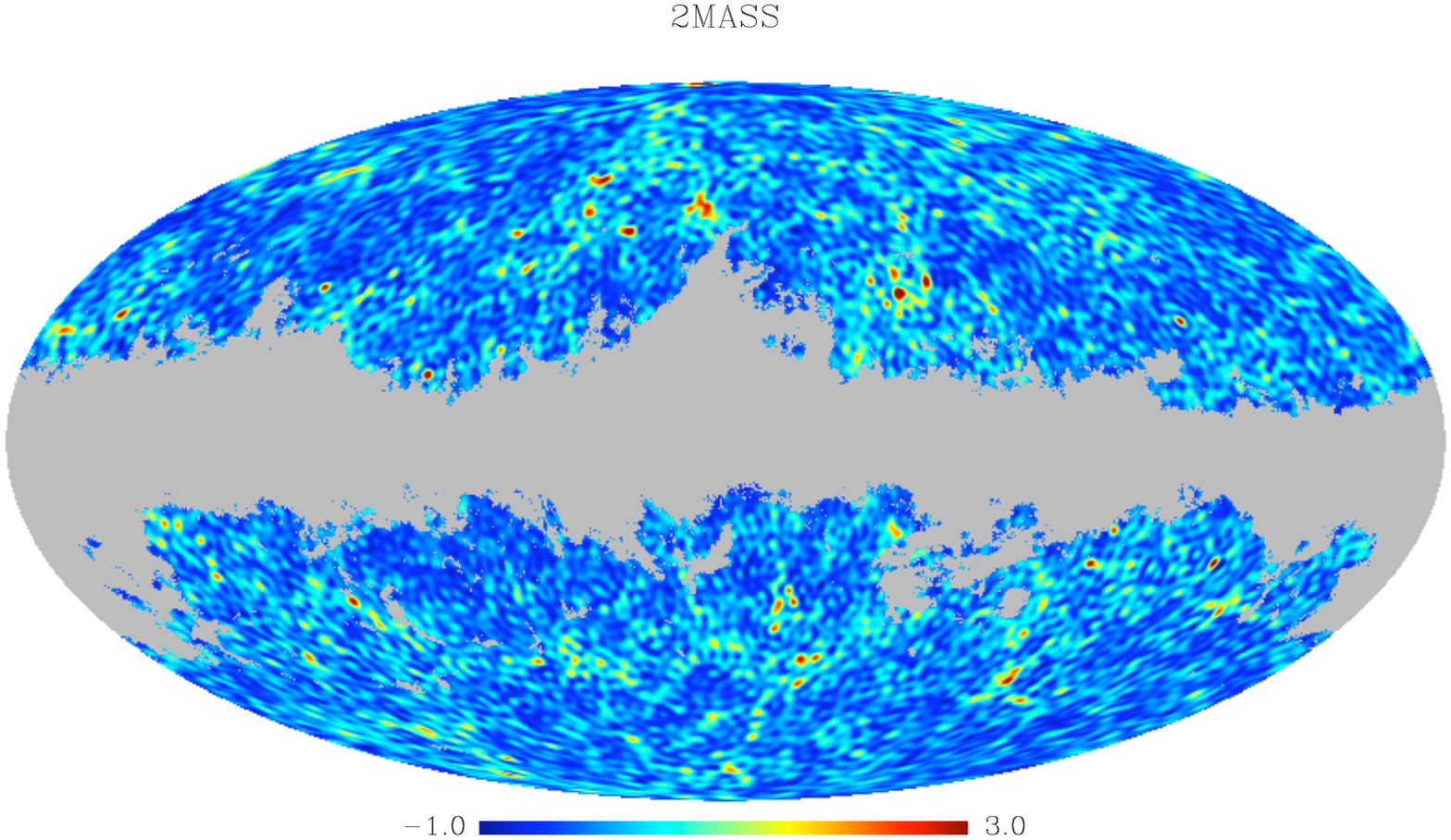}\includegraphics[width=7.5cm,angle=90]{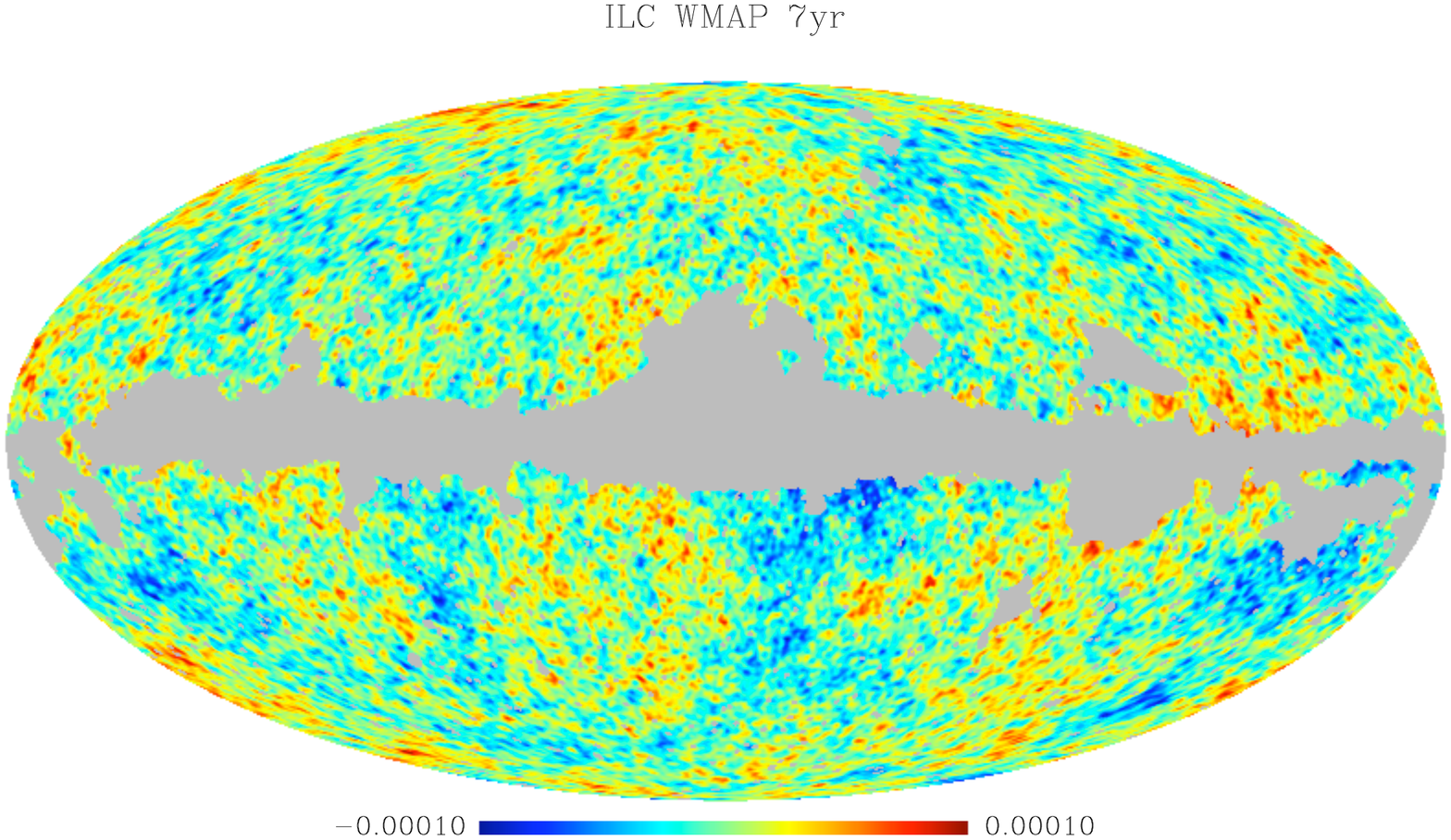}  
   \includegraphics[width=7.5cm,angle=90]{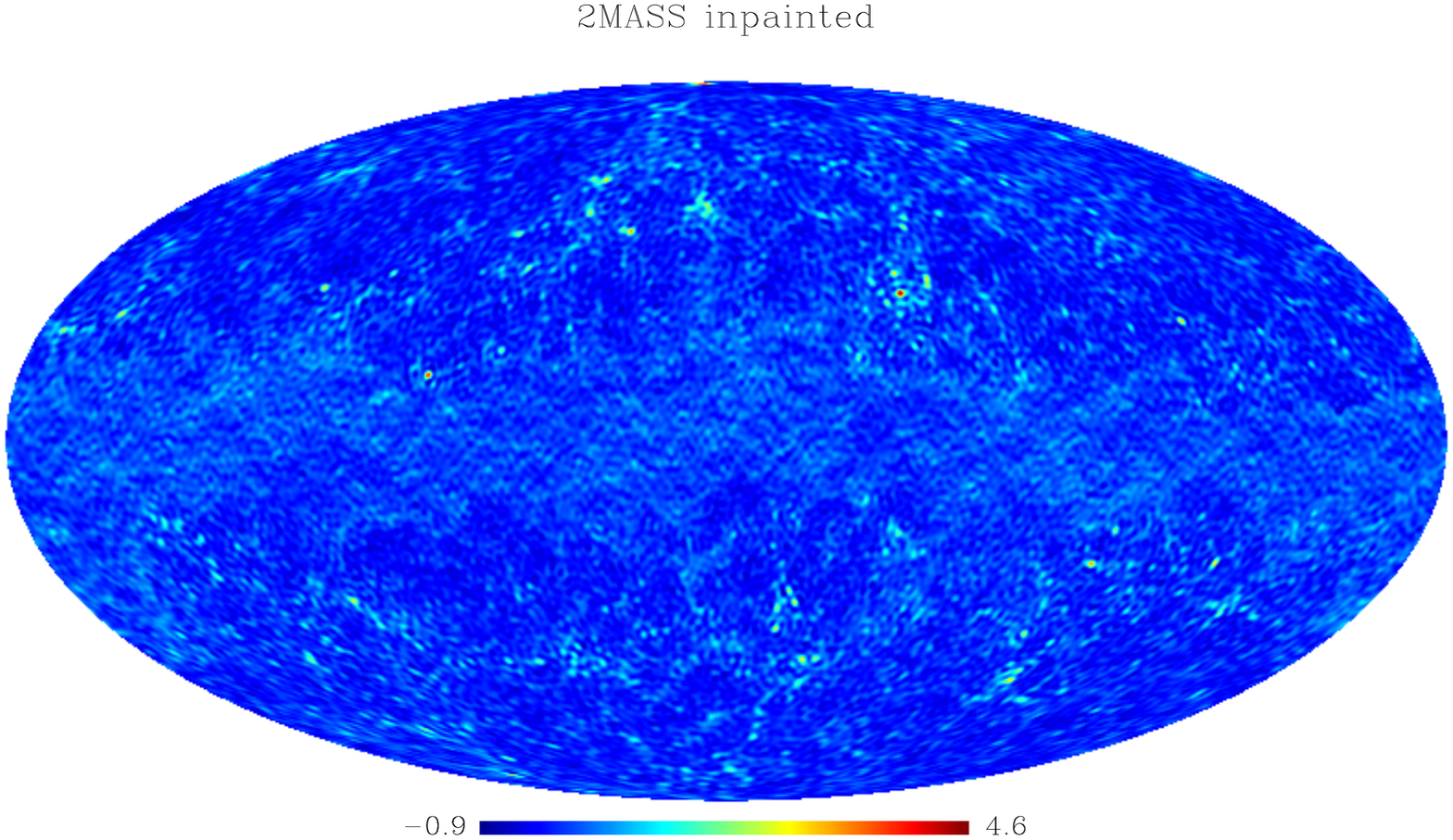}\includegraphics[width=7.5cm,angle=90]{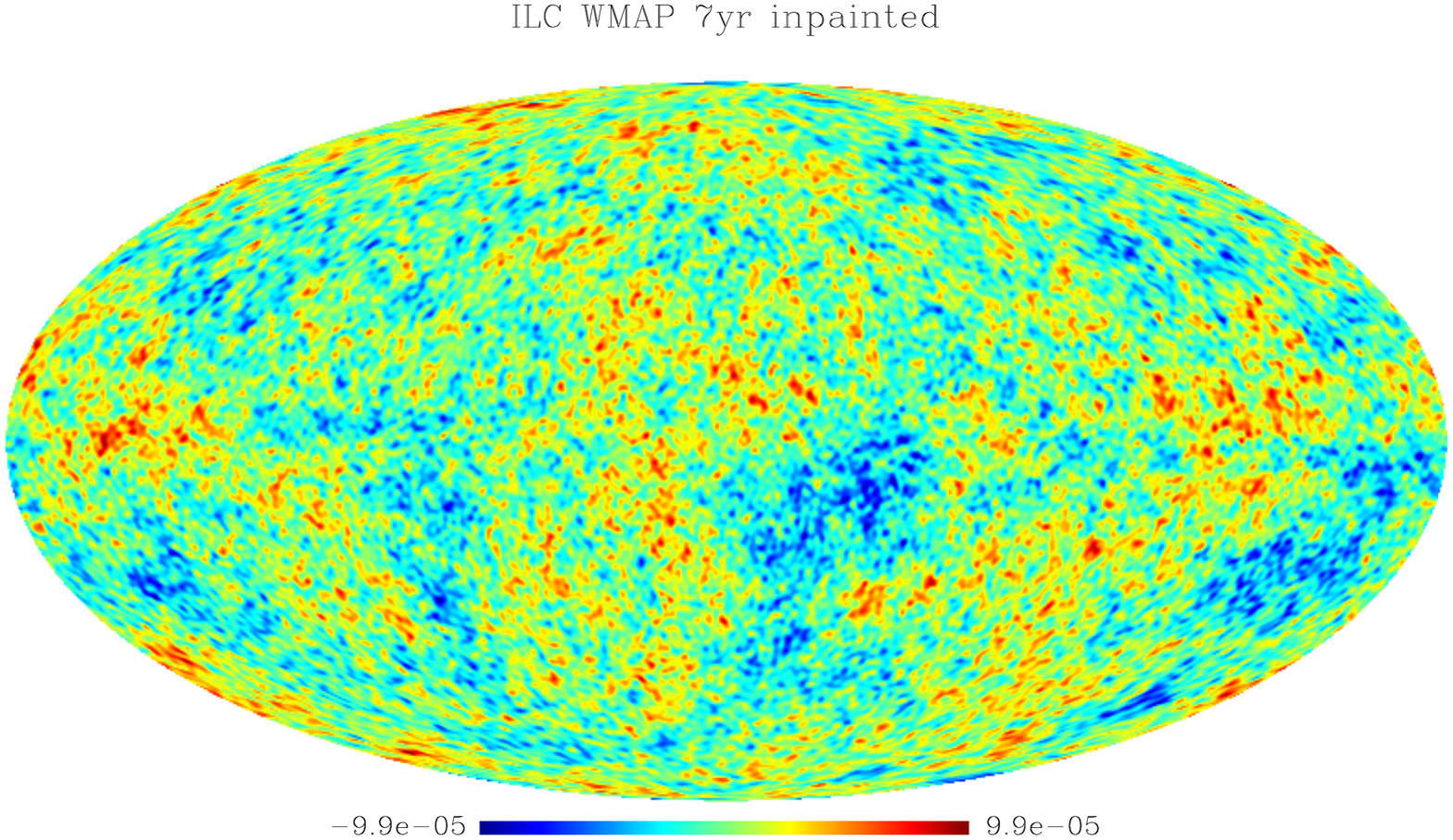} 

   \includegraphics[width=7.5cm,angle=90]{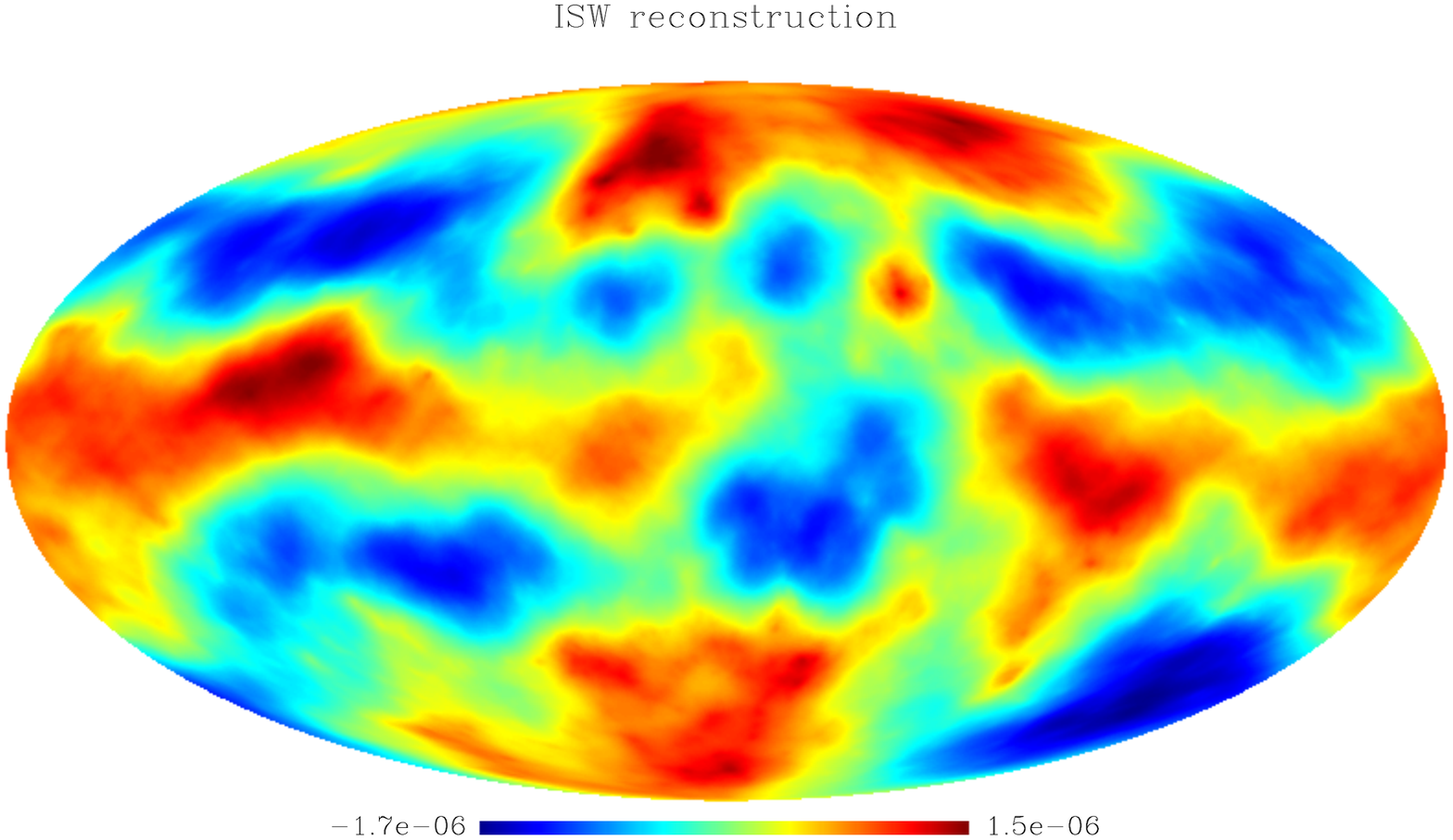} 
   \caption{Top: 2MASS map with mask (left) and WMAP 7 ILC map with mask (right). Middle: Reconstructed 2MASS (left) and
     WMAP 7 ILC (right) maps using our inpainting method.  Bottom: reconstructed ISW temperature field due to 2MASS galaxies,
     calculated using Equation \ref{eq:8}. For better visualisation of the maps as an input of the Saclay method, we
     consider only the information inside $\ell \in [2,200]$.}
  \label{fig:inpainting}
\end{figure*}

For the cosmic microwave background data, we use several maps from NASA Wilkinson Microwave
Anisotropy Probe: the internal linear combination map (ILC) for years 5 and 7 \citep[WMAP5,][]{Komatsu:2009} \citep[WMAP7,][]{Jarosik2010} and the ILC map by \cite{Delabrouille:2008} which was reconstructed using a needlets technique.  
We avoid regions which are contaminated by Galactic
emission by applying the $Kq85$ temperature mask - which roughly corresponds to the $Kp2$ mask from the third year
release (see Figure \ref{fig:inpainting}). We also substract the kinetic Doppler quadrupole contribution from the data.  WMAP simulations used to produce Tables \ref{tab:simu2mass} and \ref{tab:simuEuclid} use WMAP 7 best fit parameters for a flat $\Lambda {\rm CDM}$ universe (see
Table~\ref{tab:wmap7}).

\begin{table}[htp]
  \centering
  \begin{tabular}{|l|c||l|c||l|c|}
    \hline 
    \hline
    $\Omega_b$ & 0.0449 &
    $\Omega_m$ & 0.266 &
     $\Omega_{\Lambda}$ & 0.734 \\
     \hline $n$ & 0.963 &
     $\sigma_8$ & 0.801 &
    $h$ & 0.710 \\
    \hline $\tau$ & 0.088 &
    $w_0$ & -1.00 &
    $w_a$ & 0.0 \\
    \hline
  \end{tabular}
  \caption{Best fit WMAP 7 cosmological parameters used throughout this paper.}
  \label{tab:wmap7}
\end{table}

\begin{table}
  \centering
  \begin{tabular}{|l|c|}
    \hline
    \hline
    Method&Inpainted  \\
    &maps (with ISW)\\
    \hline
    \hline
    Simple cross-correlation  (Eq. \ref{eq:19})  & $1.30\sigma \pm 1.37$ \\
    \hline Amplitude estimation (Eq. \ref{eq:16}) & $1.00\sigma \pm 0.65$ \\
    \hline Model selection (Eq. \ref{eq:2}) & $1.00\sigma \pm 0.65$ \\
    \hline
  \end{tabular}
  \caption{Expected detections for 2MASS like survey (from lognormal results in Table~\ref{tab:simu2mass}).}
  \label{tab:exp2mass}
\end{table}

\subsection{2MASS Galaxy Survey}\label{sec:data:2mass}

The 2 Micron All-Sky Survey (2MASS) is a publicly available full-sky extended source catalogue (XSC) selected in the
near-IR \citep{Jarrett:2000me}. The near-IR selection means galaxies are surveyed deep into the Galactic plane, meaning
2MASS has a very large sky coverage, ideal for detecting the ISW signal.

Following \citep{Afshordi:2003xu}, we create a mask to exclude regions of sky where XSC is unreliable using the IR
reddening maps of \citet{Schlegel:1997yv}. Using $A_k = 0.367 \times E(B-V)$, \citet{Afshordi:2003xu} find a limit $A_K
< 0.05$ for which 2MASS is seen to 98\% complete for $K_{20} < 13.85$, where $K_{20}$ is the $K_s$-band isophotal
magnitude.  Masking areas with $A_K > 0.05$ leaves 69\% of the sky and approximately 828 000 galaxies for the analysis
(see Figure \ref{fig:inpainting}).

We use the redshift distribution computed by \cite{Afshordi:2003xu} \citep[and also used in][]{Rassat:2007KRL}, and in
order to maximise the signal, we consider one overall bin for magnitudes $12<K<14$.  The redshift distribution for 2MASS
is that shown in Figure 1 of \cite{Rassat:2007KRL} (solid black line) and peaks at $z\sim0.073$. The authors also showed
that the small angle approximation could be used for calculations relating to 2MASS, so Equations \ref{eq:cgg} and
\ref{eq:cgt} can be replaced by their simpler small angle form:
\begin{equation} C_{gT}(\ell) = \frac{-3bH_0^2\Omega_{m,0}}{c^3(\ell+1/2)^2}\int \rm{d}r \Theta D^2 H [f-1]P\left(\frac{\ell+1/2}{r}\right),\end{equation}
and
\begin{equation}C_{gg}(\ell)=b^2\int \rm{d}r \frac{\Theta^2}{r^2}D^2P\left(\frac{\ell+1/2}{r}\right)\end{equation} We
estimate the bias from the 2MASS galaxy power spectrum using the cosmology in Table \ref{tab:wmap7} and find $b=1.27 \pm
0.03$, which is lower than that found in \cite{Rassat:2007KRL}.

\subsection{Applying Sparse Inpainting to CMB and Galaxy data}
\label{sec:data:inpainting}

As discussed in Section \ref{sec:saclaymethod}, regions of missing data in galaxy and CMB maps constitute
an ill-posed problem when using a `field' based method.  We propose to use sparse inpainting \cite[see][and appendix
\ref{app:inpainting}]{starck:book10} to reconstruct the field in the regions of missing data.

We apply this method to both the WMAP7 and 2MASS maps, and the reconstructed maps are shown in Figure
\ref{fig:inpainting}.  All maps are pixelised using the HEALPix software \citep{healpix:2002,Gorski:2004by} with
resolution corresponding to NSIDE=512. The top two figures show 2MASS data (left) and ILC map (right) with the masks in
grey. The two figures in the middle show the reconstructed density fields for 2MASS data (left) and the ILC map
(right). The bottom two figures show the reconstruction of the ISW field using the inpainted 2MASS density map
and Equation \ref{eq:8} (for clarity, the first two multipoles ($\ell=0,1$) are not present in this map).

\subsection{ISW Detection using 2MASS and WMAP7 Data}
\label{sec:results}
f
We use the Saclay method (section~\ref{sec:saclaymethod}) on WMAP7 and 2MASS data with $10^6$
iterations for the Monte-Carlo iterations, $1000$ iterations for both bootstrap and nested bootstrap of the correlation
detection, and search for the ISW signal. In order to test the effect of including smaller (and possibly non-linear) scales, we perform the analysis for three different $\ell$-ranges:  $\ell \in [\ell_{\rm min},50], [\ell_{\rm min},100], [\ell_{\rm min},200]$, where $\ell_{\rm min}=2$ or $3$. We notice that inclusion or not of the quadrupole ($\ell=2$) can affect the significance level slighty, which is why we chose two values for $\ell_{\rm min}$.  We report our measured detection levels in Table
\ref{tab:resl100}. We recall the expected detections for 2MASS in Table~\ref{tab:exp2mass}, which can also be found in
the more complete set of results presented in Table \ref{tab:simu2mass}.

The results in Table \ref{tab:resl100} are compatible with the results in Table \ref{tab:simu2mass} within the errors bars.  \cite{Rassat:2007KRL} used a spectra model-comparison method over $\ell=[3-30]$, and found that the model with dark energy was marginally preferred over the null hypothesis. Over a range $\ell=[2/3-50]$, using a fields model-comparison method, we find that a model with dark energy is preferred over the null hypothesis at the $1.1-1.8 \sigma$ level, depending on the map.  More generally, our results are compatible with the earliest ISW measurement using 2MASS data \citep{Afshordi:2003xu,Rassat:2007KRL} and lie in the $1.1-2.0 \sigma$ range depending on the data and statistical test used.
  
  The simple correlation test tends to report marginally higher detection levels than the field amplitude and model comparison tests, and the model comparison test similar values as the field amplitude test, which is compatible with the predictions from the ROC curves in Figure \ref{fig:roc}.   

\begin{table*}
  \centering
  {\bf
     
 \begin{tabular}{|p{0.25\linewidth}|c||c|c|}
     \hline\hline
     Galaxy Survey & \multicolumn{3}{|c|}{2MASS} \\
     \hline CMB data & WMAP ILC 7yr & WMAP ILC 5yr & Needlets ILC 5yr\\
     \hline\hline
      Scales interval & \multicolumn{3}{|c|}{$\ell \in [2,50]$} \\
      \hline
      \hline Simple correlation detection & 1.3$\sigma$ & 1.3$\sigma$ &
1.2$\sigma$ \\
      \hline Field amplitude                     & 1.3$\sigma$ &
1.3$\sigma$ & 1.1$\sigma$ \\
      \hline Model comparison                & 1.3$\sigma$ &
1.3$\sigma$ & 1.1$\sigma$ \\
      \hline\hline
      Scales interval & \multicolumn{3}{|c|}{$\ell \in [2,100]$} \\
      \hline
      \hline Simple correlation detection & 1.5$\sigma$ & 1.4$\sigma$ &
1.3$\sigma$\\
      \hline Field amplitude                     & 1.4$\sigma$ &
1.3$\sigma$ & 1.2$\sigma$\\
      \hline Model comparison                & 1.4$\sigma$ &
1.3$\sigma$ & 1.2$\sigma$ \\
      \hline\hline
      Scales interval & \multicolumn{3}{|c|}{$\ell \in [2,200]$} \\
      \hline
      \hline Simple correlation detection & 1.4$\sigma$ & 1.4$\sigma$ &
1.3$\sigma$ \\
      \hline Field amplitude                     & 1.4$\sigma$ &
1.3$\sigma$ & 1.2$\sigma$ \\
      \hline Model comparison                & 1.4$\sigma$ &
1.3$\sigma$ & 1.2$\sigma$ \\

     \hline\hline
      Scales interval & \multicolumn{3}{|c|}{$\ell \in [3,50]$} \\
      \hline
      \hline Simple correlation detection & 2.0$\sigma$ & 1.9$\sigma$ &
1.5$\sigma$\\
      \hline Field amplitude                     & 1.8$\sigma$ &
1.7$\sigma$ & 1.4$\sigma$\\
      \hline Model comparison                & 1.8$\sigma$ &
1.7$\sigma$ & 1.4$\sigma$ \\
      \hline\hline
      Scales interval & \multicolumn{3}{|c|}{$\ell \in [3,100]$} \\
      \hline
      \hline Simple correlation detection & 2.0$\sigma$ & 2.0$\sigma$ &
1.5$\sigma$ \\
      \hline Field amplitude                     & 1.9$\sigma$ &
1.7$\sigma$ & 1.5$\sigma$ \\
      \hline Model comparison                & 1.9$\sigma$ &
1.7$\sigma$ & 1.5$\sigma$ \\
      \hline\hline
      Scales interval & \multicolumn{3}{|c|}{$\ell \in [3,200]$} \\
      \hline
      \hline Simple correlation detection & 1.9$\sigma$ & 1.9$\sigma$ &
1.6$\sigma$ \\
      \hline Field amplitude                     & 1.8$\sigma$ &
1.7$\sigma$ & 1.5$\sigma$ \\
      \hline Model comparison                & 1.8$\sigma$ &
1.7$\sigma$ & 1.5$\sigma$ \\
      \hline

   \end{tabular}  }
  \caption{Detection levels obtained with different CMB maps and different scales intervals (considering with and without quadrupole) 
    using the Saclay method described in \ref{sec:saclaymethod}.} 
  \label{tab:resl100}
\end{table*}

\section{Discussion}\label{sec:discussion}

In this paper we have extensively reviewed the numerous methods in the literature which are used to detect and measure
the presence of an ISW signal using maps for the CMB and local tracers of mass. We noticed that the variety of methods
used can lead to different and conflicting conclusions. We also noted two broad classes of methods: one which uses the
cross-correlation spectrum as the measure and the other which uses the reconstructed ISW temperature field.

We identified the advantages and disadvantages of all methods used in the literature and concluded that: 
\begin{enumerate}
\item Using the fields (instead of spectra) as input required \emph{only} the primordial CMB to be Gaussian. This requires reconstruction of the ISW field, which is difficult with missing data.
\item The ill-posed problem of missing data can be solved using sparse inpainting, a method which does not introduce spurious correlations between maps.
\item Assuming the estimator was Gaussian led to under-estimation of the signal estimation.
\item A series of statistical tests could provide complementary information.
\end{enumerate}

This led us to construct a new and complete method for detecting and measuring the ISW effect.  The method is summarised as follows: 
\begin{enumerate}
\item Apply sparse inpainting to both the galaxy and CMB maps which may have different masks, essentially reconstructing missing data around the Galactic plane and bulge. 
\item Test for simple correlation, using a double bootstrap (one to estimate the variance and another to estimate the confidence) and using the fields as input. This returns a model-independent detection level.
\item Reconstruct the ISW signal using expected cosmology and the inpainted galaxy density map;
\item Estimate the signal amplitude using the fields as input, and apply a bootstrap (or MC) to the estimator. This validates both the signal and the model.
\item Apply the fields model comparison test, essentially testing whether the data prefers a given model over the null hypothesis, and measure relative fit of models.
\end{enumerate}

The method we present in this paper makes only one assumption: that the primordial CMB temperature field behaves like a Gaussian random field.  The method is general in that it `allows' the galaxy field to behave as a lognormal field, but does not automatically assume that the galaxy field is lognormal.

We first applied our method to 2MASS and Euclid simulations. We find that it is difficult to detect the ISW significantly using 2MASS simulations, and find no difference between assuming the underlying galaxy field is Gaussian or lognormal, and only mild differences depending on the statistical test used. With a Euclid-like survey, we expect high detection levels, even with incomplete sky coverage - we expect $\sim 7 \sigma$ detection level using the simple correlation method, and $\sim 4.7\sigma$ detection level using the fields amplitude or method comparison techniques.  These detections levels are the same whether the Euclid galaxy field follows a Gaussian or lognormal distribution. Our results also show that the inpainting method does not introduce spurious correlations between maps.

We applied this method to WMAP7 and 2MASS data, and found that our results were
comparable with early detections of the ISW signal using 2MASS data \citep{Afshordi:2003xu,Rassat:2007KRL} and lied
roughly in the $1.1-2.0\sigma$ range. These results are also compatible with the simulations we ran for the 2MASS survey.
  
The last test we performed, the model comparison test, asks the much more pertinent question of whether the data prefers
a $\Lambda$CDM model to the null hypothesis (i.e. no curvature and no dark energy). Using this test, we find a $1.1-1.8\sigma$ detection for ranges $\ell=[2/3-50]$ and $1.2-1.9\sigma$ for ranges $\ell=[2/3-100/200]$, which is
sometimes higher than what was previously reported in \cite{Rassat:2007KRL} using a spectra models comparison
  test, without sparse inpainting or bootstrapping.  A by-product of this measurement is the reconstruction of the
  temperature ISW field due to 2MASS galaxies, reconstructed with full sky coverage.

By applying our method to different estimations of CMB maps, we highlight the importance of component
  separation on the ISW detection.  Table~\ref{tab:resl100} shows scores between $1.1-2.0\sigma$ on different WMAP maps. We were also able to detect the ISW signal at $2.7\sigma$ using a another map (not presented in this paper). The influence of the component separation method on the quality of the estimation needs to be more deeply understood in future works.

\begin{acknowledgements}
  This work has been supported by the European Research Council grant SparseAstro (ERC-228261). We thank Niayesh Afshordi, Ofer Lahav,
  John Peacock, Alexandre R\'efr\'egier and the anonymous referee for useful discussions. We also thank Jacques Delabrouille for providing the needlets ILC 5yr map from the WMAP 5yr data.  We used iCosmo \footnote{\url{http://www.icosmo.org}} software \citep{Refregier:2011}, Healpix software \citep{healpix:2002,Gorski:2004by},
  ISAP\footnote{\url{http://jstarck.free.fr/isap.html}} software, the 2MASS catalogue
  \footnote{\url{http://www.ipac.caltech.edu/2mass/}}, the \emph{WMAP} data \footnote{\url{http://map.gsfc.nasa.gov}}
  and the Galaxy extinction maps of \cite{Schlegel:1997yv}.
\end{acknowledgements}
\appendix

\section{Sparse Inpainting}
\label{app:inpainting}
The goal of inpainting is to restore missing or damaged regions of an image, in such a way that the restored map has the same
statistical properties as the underlying unmasked map \citep{Elad2005}.  Sparse Inpainting has been proposed for filling
the gaps in CMB maps \citep{inpainting:abrial06,Abrial2008} and for weak lensing mass map reconstruction
\citep{starck:pires08,pires10}. In \cite{perotto10}, it has been shown that the sparse inpainting method does not
destroy the CMB weak lensing signal, and is therefore an elegant way to handle the mask problem.

The inpainting problem can be defined as follows.  Let $X$ be the ideal complete image, $Y$ the observed incomplete
image (images can be fields on the sphere) and $L$ the binary mask (i.e. $L[k,l] = 1$ if we have information at pixel
$(k,l)$, $L[k,l] = 0$ otherwise). In short, we have: $Y = L X$.

Inpainting consists in recovering $X$ knowing $Y$ and $L$.  The masking effect can be thought of as a loss of sparsity
in the spherical harmonic domain since the information required to define the map has been spread across the spherical
harmonic basis.

Sparsity means that most of the information is concentrated in a few coefficients, which when sorted from the largest to
the smallest, follow an exponential decay. More details can be found in \citep{starck:book10}.  In this paper, the
chosen `dictionary' is the spherical harmonic domain.

Denoting the spherical harmonic basis as $\Phi$ (so $\Phi^T$ is the spherical harmonic transform, i.e. the projector
onto the spherical harmonic space), $|| z ||_0$ the $l_0$ pseudo-norm, i.e. the number of non-zero entries in $z$ and
$|| z ||$ the classical $l_2$ norm (i.e. $ || z ||^2 = \sum_k (z_k)2 $), we thus want to minimise:
\begin{equation}
\min_X  \| \Phi^T X \|_0 \quad   \text{subject to}  \quad   \parallel Y - L X  \parallel_{\ell_2} \le \sigma,
\label{minimisation}
\end{equation}
where $\sigma$ stands for the noise standard deviation in the noisy case.  It has also been shown that if $X$ is sparse
enough, the $l_0$ pseudo-norm can also be replaced by the convex $l_1$ norm (i.e. $ || z ||_1 = \sum_k | z_k | $)
\citep{cur:donoho_01b}. The solution of such an optimisation task can be obtained through an iterative thresholding
algorithm called MCA \citep{inpainting:elad05,starck:jalal06,starck:book10} :
\begin{equation}
   X^{n+1} = \Delta_{\Phi,\lambda_n}(X^{n} + Y - L X^n)
\label{eqn_mca}
\end{equation}
where the nonlinear operator $\Delta_{\Phi,\lambda}(Z)$ consists in:
\begin{enumerate}
\item decomposing the signal $Z$ on the dictionary $\Phi$ to derive the coefficients $\alpha = \Phi^T Z$ (i.e. $\alpha$
  is the vector containing the $a_{\ell m}$ coefficients).
\item threshold the coefficients: ${\tilde \alpha} = \rho(\alpha, \lambda)$, 
where the thresholding operator $\rho$   can either
be a hard thresholding (i.e. $\rho(\alpha_i, \lambda) = \alpha_i$ if $ | \alpha_i | > \lambda$ and $0$ otherwise)
 or a soft thresholding (i.e.

  $\rho(\alpha_i, \lambda) = \mathrm{sign}(\alpha_i) \mathrm{max}(0, | \alpha_i |  - \lambda)$). 
 The hard thresholding corresponds to the $l_0$ 
 optimisation problem while the soft-threshold solves that for $l_1$.
\item reconstruct $\tilde Z$ from the thresholds coefficients ${\tilde \alpha}$.
\end{enumerate}
The threshold parameter $\lambda_n$ decreases with the iteration number and it plays a role similar to the cooling parameter  
of the simulated annealing techniques, i.e. it allows the solution to escape from local minima. 

It has been shown in~\cite{Abrial2008} that this inpainting technique leads to accurate CMB recovery results. 
More details relative to this optimisation problem can be found in \cite{CombettesWajs05,starck:book10} and theoretical justifications
for CMB sparse inpainting in \cite{rauhut10}.
The software is available in the Multi-Resolution on the Sphere (MRS) package \footnote{http://jstarck.free.fr/mrs.html}.  

\section{Bootstrapping detection tests} \label{app:bootstrap}

Most detection methods rely on a z-score whose confidence is computed assuming a Gaussian distribution (see
section~\ref{sec:method:review}). This assumption is valid within the central limit theorem and so is not always
true. Using an estimator of the true distribution will give more reliable confidence levels. To estimate the true
distribution, two approaches are generally used: Monte-Carlo simulations when a model and its parameters are available
or non-parametric bootstraps methods.

In order to build a fully non-parametric and cosmology independent correlation detection test, we propose to use the
wild bootstrap in order to estimate both estimator variance and confidence interval. The remaining question will be the
reliability of such test when bootstraps are sometimes known to underestimate confidence intervals. The answer to the
question depends mainly on the choice of domain (e.g. pixels, spherical harmonics) and the method chosen for estimating.

Working with correlated data is difficult and estimated values are generally different to the true ones. We chose to
work in the spherical harmonic domain, which produces heteroskedastic (i.e., the power spectrum is scale dependent) and
uncorrelated data. For heteroskedastic and uncorrelated data, bootstraps methods like wild bootstrap can be
used. \cite{Flachaire2005,Davidson2008} showed the good behaviour of wild bootstrap in many situations.  We propose here
a short introduction to this method, which we use to estimate $p$-values and estimator variances.

\subsection{Wild Bootstrap and Regression Model} \label{app:bootstrap:correlation}

Bootstrap methods were introduced by \cite{Efron1979} as a generalisation of the jackknife and its concept has been
extended to many situations. For the regression case with heteroskedastic data, wild bootstrap was developed by
\cite{Liu1988}, who established the ability of wild bootstrap to provide refinements for the linear regression model
with heteroskedastic disturbances. \cite{Mammen1993} showed that the wild bootstrap was asymptotically justified, in the
sense that the asymptotic distribution of various statistics is the same as the asymptotic distribution of their wild
bootstrap counterparts.

If you look at the correlation estimator~(Equation \ref{eq:9}) and the amplitude estimation~(Equation \ref{eq:1}), they both suppose an
underlying linear model:
\begin{equation}
  \label{eq:21}
    \delta^{O}_{\ell m} = \rho \delta^{C}_{\ell m} + \sigma_{\ell m}~,
\end{equation}
where $\delta^{O}$ is the observed field which contains the signal of interest (e.g. CMB), $\delta^{C}$ a tracer of the
signal of interest (e.g. matter density, ISW reconstruction), $\rho$ is either the correlation coefficient and the
amplitude of the signal (depending of the estimator) and $\sigma$ is the noise which is dependent of the position.

The originality of the wild bootstrap relies in the fact that its data generating process (DGP) creates new samples
  without any prior on the distribution of the noise. This process uses the residual, i.e. $u_{\ell m} =
\delta^{O}_{\ell m} - \hat{\rho} \delta^{C}_{\ell m} $, where $\hat{\rho}$ is the parameter estimate.  Then, new samples
are created using the following formula,
\begin{align}
  \label{eq:22}
  \delta^{O\star}_{\ell m} &= \hat{\rho} \delta^{C}_{\ell m} + u^{\star}_{\ell m}~, \\
  u^{\star}_{\ell m} &= \frac{u_{\ell m} \varepsilon}{1 - h_{\ell m}}, \quad\\
  h_{\ell m} &= \frac{\abs{\delta^{C}_{\ell m}}^2}{\sum_{\ell,m} \abs{\delta^{C}_{\ell m}}^2 }
\end{align}
where ${ }^*$ indicate the bootstrapped version of the sample and $\varepsilon$ is a random variable following a given
distribution. The performance of the wild bootstrap mainly depends on the distribution of $\varepsilon$, which must
verify some conditions like having a null mean and a unit variance.

For example \cite{Liu1988} proposed to use Rademacher variables for the distribution,
\begin{equation}
  \label{eq:20}
  \varepsilon = \begin{cases}
    1 & \text{with probability } 0.5~, \\
    -1 & \text{with probability } 0.5~. \\
    \end{cases}
  \end{equation}
  \cite{Davidson2008} showed that exact inference (up to a theoretical accuracy) is possible if $\delta^C_{\ell m}$
    is  independent from all the disturbance $u_{\ell m}$ and if the test distribution is symmetric about 0.
    Moreover, the rate of convergence of the error in the rejection probabilities (ERP) is at most $n^{-3/2}$ with symmetric errors and $n^{-1/2}$ with
  asymmetric errors ($n$ is the size of the observed sample). The ERP is the difference between the actual rejection
  frequency under the null hypothesis and the level of the test (e.g. about 0.045 for a $2\sigma$ detection). In other
  words, this is the precision error on the inferred $p$-value. Notice that it mainly depends on the sample size.

\subsection{Parametric bootstrap}
\label{sec:parametric-bootstrap}

The wild bootstrap is very useful when little is known about the noise. However, sometimes the noise distribution is
well known and its parameters can be estimated from the observations. In this case, it is possible to use this
distribution in order to generate new samples. For example, if we model the CMB primordial distribution by a multivariate
normal law, the $u^\star_{\ell m}$ can be generated as, $u^\star_{\ell m} \sim \mathcal{N}(0,C_{\mathrm{CMB}})$
where $C_{\mathrm{CMB}}$ is the power spectra of the primordial CMB. Notice, that using the theoretical CMB power
spectra leads to a Monte-Carlo methods which is then dependent on the chosen cosmology.

\subsection{Estimating the estimator variance}
\label{sec:estim-estim-vari}

When the variance of an estimator is unknown, a bootstrap can be used to estimate it with
the following scheme:
\begin{enumerate}
\item Estimate the parameter vector $\hat{\rho}$ from the observed data;
\item Create $B$ bootstrap samples, using Equation \ref{eq:22} or another method (see
  section~\ref{sec:parametric-bootstrap});
\item For each bootstrap sample,estimate the parameter vector $\rho^{\star}_i$, $i\in [1,B]$,
\item Compute the estimation of the variance of the estimator,
\begin{equation}
    \label{eq:11}
     \mathrm{Var}(\hat{\rho}) = \frac{1}{B-1} \sum^B_{i=1} (\rho^{\star}_i - \bar{\rho}^{\star})^2~, \quad
      \bar{\rho}^{\star} = \frac{1}{B} \sum^B_{i=1} \rho^{\star}_i~.
 \end{equation}
\end{enumerate}

\subsection{Estimating the $p$-value of a test}
\label{sec:estimating-pvalue}

The estimation of the confidence interval of a hypothesis test requires the generation of samples under the null
hypothesis. Inside the wild bootstrap DGP, that leads to set $\hat{\rho} = 0$ when using Equation \ref{eq:22} for new
samples (it also remain true with a parametric bootstrap). Using this property, the scheme for computing the $p$-value
is the following,
\begin{enumerate}
\item Estimate the parameter vector and its variance (e.g. using the previous scheme in
  \ref{sec:estim-estim-vari}), $\hat{\rho}$ and $\mathrm{Var}(\hat{\rho})$;
\item Compute the z-score,
  \begin{equation}
    \label{eq:17}
    \hat{\tau} = \hat{\rho} / \sqrt{\mathrm{Var}(\hat{\rho})}~;
  \end{equation}
\item Create $B$ bootstrap samples which follow the null hypothesis using \ref{eq:22} (or
    section~\ref{sec:parametric-bootstrap}) and $\hat{\rho} = 0$ ;
\item For each bootstrap sample, estimate the parameter, $\rho^{\star}_i$, and its variance
  $\mathrm{Var}(\rho^{\star}_i)$, $i\in [1,B]$;
\item For each bootstrap sample, compute the correspondant z-score, 
  \begin{equation}
    \label{eq:18}
    \tau^{\star}_i = \rho^{\star}_i / \sqrt{\mathrm{Var}(\rho^{\star}_i)}~;
  \end{equation}
\item Compute the bootstrapped $p$-value,
\begin{equation}
    \label{eq:12}
    p^{\star} = \frac{1}{B} \sum_i \boldsymbol{I}(\tau^{\star}_i > \hat{\tau})~,
  \end{equation}
  where $\boldsymbol{I}$ denotes the indicator function, which is equal to 1 when its
  argument is true and 0 otherwise.
\end{enumerate}

Equation \ref{eq:12} can be a little too rough and may lead to inaccurate results when the answer is just between
two bootstrap samples. One way to reduce this effect is by smoothing the empirical distribution \citep{Racine2007}, for
example with a Gaussian kernel,
\begin{equation}
  p^{\star} = 1 - \frac{1}{B} \sum_{i=0}^{B-1}{\varphi} \left(\frac{\hat{\tau} - \tau^{\star}_i}{h}\right),
   \label{eq:13}
\end{equation}
where $\varphi$ is the Gaussian cumulative distribution function and $h$ the bandwidth of the kernel \cite[chosen
depending on the required precision, see][]{Racine2007}.

\subsection{Reliability of bootstrap}
\label{sec:reli-bootstr}

One can be perplexed by the efficiency and the reliability of the bootstrap method. Many studies have proved that these
  methods outperform asymptotic tests \citep[i.e. using the asymptotic distribution, see][]{Davidson2006} and sometimes
  perform almost as well as exact tests \citep[ (i.e. with the true distribution, we refer the reader to][for more information]{Efron1987,DiCiccio1996,Hall1995}. The efficiency is mostly dependent on three key
  elements: 1) the Data Generating Process (DGP) under both hypotheses (often generating under the alternative
  hypothesis is non trivial), 2) the number of observations and 3) the number of bootstrapped samples. The variance on $p$ is given by $p(1-100)/B$, where $B$ is the number of bootstraps. While the number
  of bootstrapped sample can be easily tackled, the DGP on the chosen method and the number of observation depends on
  many parameters (e.g. configuration, instruments, acquisition time). Even with the most efficient DGP, the precision
  of the bootstrap methods is limited by the number of observations, often in $\mathcal{O}(1/\sqrt{n})$, sometimes
  $\mathcal{O}(1/n)$. In other words, bootstrap methods are mostly inaccurate when the observed value is at the tail of
  the involved distribution (i.e. and so rare event). For example, a $p$-value estimated using $100$ observations will
  be $\pm 0.1$ and be inefficient for characterizing high confidence, but can still be useful for rejecting.

\bibliographystyle{aa}
\bibliography{./references}

\begin{thebibliography}{98}
\expandafter\ifx\csname natexlab\endcsname\relax\def\natexlab#1{#1}\fi

\bibitem[{{Abrial} {et~al.}(2007){Abrial}, {Moudden}, {Starck}, {Bobin},
  {Fadili}, {Afeyan}, \& {Nguyen}}]{inpainting:abrial06}
{Abrial}, P., {Moudden}, Y., {Starck}, J., {et~al.} 2007, \jfaa, 13, 729--748

\bibitem[{{Abrial} {et~al.}(2008){Abrial}, {Moudden}, {Starck}, {Fadili},
  {Delabrouille}, \& {Nguyen}}]{Abrial2008}
{Abrial}, P., {Moudden}, Y., {Starck}, J.-L., {et~al.} 2008, Statistical
  Methodology, 5, 289

\bibitem[{{Adelman-McCarthy} {et~al.}(2008){Adelman-McCarthy}, {Ag{\"u}eros},
  {Allam}, {Allende Prieto}, {Anderson}, {Anderson}, {Annis}, {Bahcall},
  {Bailer-Jones}, {Baldry}, {Barentine}, {Bassett}, {Becker}, {Beers}, {Bell},
  {Berlind}, {Bernardi}, {Blanton}, {Bochanski}, {Boroski}, {Brinchmann},
  {Brinkmann}, {Brunner}, {Budav{\'a}ri}, {Carliles}, {Carr}, {Castander},
  {Cinabro}, {Cool}, {Covey}, {Csabai}, {Cunha}, {Davenport}, {Dilday}, {Doi},
  {Eisenstein}, {Evans}, {Fan}, {Finkbeiner}, {Friedman}, {Frieman},
  {Fukugita}, {G{\"a}nsicke}, {Gates}, {Gillespie}, {Glazebrook}, {Gray},
  {Grebel}, {Gunn}, {Gurbani}, {Hall}, {Harding}, {Harvanek}, {Hawley},
  {Hayes}, {Heckman}, {Hendry}, {Hindsley}, {Hirata}, {Hogan}, {Hogg}, {Hyde},
  {Ichikawa}, {Ivezi{\'c}}, {Jester}, {Johnson}, {Jorgensen}, {Juri{\'c}},
  {Kent}, {Kessler}, {Kleinman}, {Knapp}, {Kron}, {Krzesinski}, {Kuropatkin},
  {Lamb}, {Lampeitl}, {Lebedeva}, {Lee}, {Leger}, {L{\'e}pine}, {Lima}, {Lin},
  {Long}, {Loomis}, {Loveday}, {Lupton}, {Malanushenko}, {Malanushenko},
  {Mandelbaum}, {Margon}, {Marriner}, {Mart{\'{\i}}nez-Delgado}, {Matsubara},
  {McGehee}, {McKay}, {Meiksin}, {Morrison}, {Munn}, {Nakajima}, {Neilsen},
  {Newberg}, {Nichol}, {Nicinski}, {Nieto-Santisteban}, {Nitta}, {Okamura},
  {Owen}, {Oyaizu}, {Padmanabhan}, {Pan}, {Park}, {Peoples}, {Pier}, {Pope},
  {Purger}, {Raddick}, {Re Fiorentin}, {Richards}, {Richmond}, {Riess}, {Rix},
  {Rockosi}, {Sako}, {Schlegel}, {Schneider}, {Schreiber}, {Schwope}, {Seljak},
  {Sesar}, {Sheldon}, {Shimasaku}, {Sivarani}, {Smith}, {Snedden}, {Steinmetz},
  {Strauss}, {SubbaRao}, {Suto}, {Szalay}, {Szapudi}, {Szkody}, {Tegmark},
  {Thakar}, {Tremonti}, {Tucker}, {Uomoto}, {Vanden Berk}, {Vandenberg},
  {Vidrih}, {Vogeley}, {Voges}, {Vogt}, {Wadadekar}, {Weinberg}, {West},
  {White}, {Wilhite}, {Yanny}, {Yocum}, {York}, {Zehavi}, \&
  {Zucker}}]{SDSS:2008}
{Adelman-McCarthy}, J.~K., {Ag{\"u}eros}, M.~A., {Allam}, S.~S., {et~al.} 2008,
  \apjs, 175, 297

\bibitem[{Afshordi {et~al.}(2004)Afshordi, Loh, \& Strauss}]{Afshordi:2003xu}
Afshordi, N., Loh, Y.-S., \& Strauss, M.~A. 2004, Phys. Rev., D69, 083524

\bibitem[{{Afshordi} {et~al.}(2011){Afshordi}, {Slosar}, \&
  {Wang}}]{Afshordi:2011}
{Afshordi}, N., {Slosar}, A., \& {Wang}, Y. 2011, \jcap, 1, 19

\bibitem[{{Ag{\"u}eros} {et~al.}(2006){Ag{\"u}eros}, {Anderson}, {Margon},
  {Posselt}, {Haberl}, {Voges}, {Annis}, {Schneider}, \&
  {Brinkmann}}]{SDSS:optical2006}
{Ag{\"u}eros}, M.~A., {Anderson}, S.~F., {Margon}, B., {et~al.} 2006, \aj, 131,
  1740

\bibitem[{{Albrecht} {et~al.}(2006){Albrecht}, {Bernstein}, {Cahn}, {Freedman},
  {Hewitt}, {Hu}, {Huth}, {Kamionkowski}, {Kolb}, {Knox}, {Mather}, {Staggs},
  \& {Suntzeff}}]{DETF}
{Albrecht}, A., {Bernstein}, G., {Cahn}, R., {et~al.} 2006, ArXiv Astrophysics
  e-prints

\bibitem[{{Amara} \& {R{\'e}fr{\'e}gier}(2007)}]{Amara:2007}
{Amara}, A. \& {R{\'e}fr{\'e}gier}, A. 2007, \mnras, 381, 1018

\bibitem[{{Anderson} {et~al.}(2001){Anderson}, {Fan}, {Richards}, {Schneider},
  {Strauss}, {Vanden Berk}, {Gunn}, {Knapp}, {Schlegel}, {Voges}, {Yanny},
  {Bahcall}, {Bernardi}, {Brinkmann}, {Brunner}, {Csab{\'a}i}, {Doi},
  {Fukugita}, {Hennessy}, {Ivezi{\'c}}, {Kunszt}, {Lamb}, {Loveday}, {Lupton},
  {McKay}, {Munn}, {Nichol}, {Szokoly}, \& {York}}]{SDSS:qso2001}
{Anderson}, S.~F., {Fan}, X., {Richards}, G.~T., {et~al.} 2001, \aj, 122, 503

\bibitem[{{Bassett} \& {Afshordi}(2010)}]{Bassett2010}
{Bassett}, B.~A. \& {Afshordi}, N. 2010, ArXiv e-prints

\bibitem[{{Bennett} {et~al.}(1990){Bennett}, {Smoot}, \& {Kogut}}]{COBE:1990}
{Bennett}, C.~L., {Smoot}, G.~F., \& {Kogut}, A. 1990, in Bulletin of the
  American Astronomical Society, Vol.~22, Bulletin of the American Astronomical
  Society, 1336--+

\bibitem[{{Bernardeau} {et~al.}(2002){Bernardeau}, {Colombi}, {Gazta{\~n}aga},
  \& {Scoccimarro}}]{Bernardeau:2002}
{Bernardeau}, F., {Colombi}, S., {Gazta{\~n}aga}, E., \& {Scoccimarro}, R.
  2002, \physrep, 367, 1

\bibitem[{{Boldt}(1987)}]{Boldt1987}
{Boldt}, E. 1987, \physrep, 146, 215

\bibitem[{Boughn \& Crittenden(2004)}]{Boughn:2003yz}
Boughn, S. \& Crittenden, R. 2004, Nature, 427, 45

\bibitem[{{Boughn} \& {Crittenden}(2002)}]{Boughn:2002}
{Boughn}, S.~P. \& {Crittenden}, R.~G. 2002, Physical Review Letters, 88,
  021302

\bibitem[{Boughn \& Crittenden(2005)}]{Boughn:2004zm}
Boughn, S.~P. \& Crittenden, R.~G. 2005, New Astron. Rev., 49, 75

\bibitem[{{Boughn} {et~al.}(1998){Boughn}, {Crittenden}, \&
  {Turok}}]{Boughn:1998}
{Boughn}, S.~P., {Crittenden}, R.~G., \& {Turok}, N.~G. 1998, \na, 3, 275

\bibitem[{{Cabr{\'e}} {et~al.}(2007){Cabr{\'e}}, {Fosalba}, {Gazta{\~n}aga}, \&
  {Manera}}]{Cabre:2007}
{Cabr{\'e}}, A., {Fosalba}, P., {Gazta{\~n}aga}, E., \& {Manera}, M. 2007,
  \mnras, 381, 1347

\bibitem[{{Cabr{\'e}} {et~al.}(2006){Cabr{\'e}}, {Gazta{\~n}aga}, {Manera},
  {Fosalba}, \& {Castander}}]{Cabre:2006qm}
{Cabr{\'e}}, A., {Gazta{\~n}aga}, E., {Manera}, M., {Fosalba}, P., \&
  {Castander}, F. 2006, \mnras, 372, L23

\bibitem[{{Cai} {et~al.}(2010){Cai}, {Cole}, {Jenkins}, \& {Frenk}}]{Cai:2010}
{Cai}, Y., {Cole}, S., {Jenkins}, A., \& {Frenk}, C.~S. 2010, \mnras, 407, 201

\bibitem[{Carroll {et~al.}(2005)}]{Carroll:2004de}
Carroll, S.~M. {et~al.} 2005, Phys. Rev., D71, 063513

\bibitem[{{Cole} {et~al.}(2001){Cole}, {Norberg}, {Baugh}, {Frenk},
  {Bland-Hawthorn}, {Bridges}, {Cannon}, {Colless}, {Collins}, {Couch},
  {Cross}, {Dalton}, {De Propris}, {Driver}, {Efstathiou}, {Ellis},
  {Glazebrook}, {Jackson}, {Lahav}, {Lewis}, {Lumsden}, {Maddox}, {Madgwick},
  {Peacock}, {Peterson}, {Sutherland}, \& {Taylor}}]{Cole:2001}
{Cole}, S., {Norberg}, P., {Baugh}, C.~M., {et~al.} 2001, \mnras, 326, 255

\bibitem[{Combettes \& Wajs(2005)}]{CombettesWajs05}
Combettes, P.~L. \& Wajs, V.~R. 2005, \siammms, 4, 1168--1200

\bibitem[{{Condon} {et~al.}(1998){Condon}, {Cotton}, {Greisen}, {Yin},
  {Perley}, {Taylor}, \& {Broderick}}]{Condon:1998}
{Condon}, J.~J., {Cotton}, W.~D., {Greisen}, E.~W., {et~al.} 1998, \aj, 115,
  1693

\bibitem[{{Corasaniti} {et~al.}(2005){Corasaniti}, {Giannantonio}, \&
  {Melchiorri}}]{Corasaniti:2005}
{Corasaniti}, P., {Giannantonio}, T., \& {Melchiorri}, A. 2005, \prd, 71,
  123521

\bibitem[{Crittenden \& Turok(1996)}]{Crittenden:1995ak}
Crittenden, R.~G. \& Turok, N. 1996, Phys. Rev. Lett., 76, 575

\bibitem[{Davidson \& MacKinnon(2006)}]{Davidson2006}
Davidson, A. \& MacKinnon, J.~G. 2006, Journal of Econometrics, 133, 421

\bibitem[{Davidson \& Flachaire(2008)}]{Davidson2008}
Davidson, R. \& Flachaire, E. 2008, Journal of Econometrics, 146, 162

\bibitem[{{Delabrouille} {et~al.}(2009){Delabrouille}, {Cardoso}, {Le Jeune},
  {Betoule}, {Fay}, \& {Guilloux}}]{Delabrouille:2008}
{Delabrouille}, J., {Cardoso}, J., {Le Jeune}, M., {et~al.} 2009, \aap, 493,
  835

\bibitem[{DiCiccio \& Efron(1996)}]{DiCiccio1996}
DiCiccio, T.~J. \& Efron, B. 1996, Statistical Science, 11, 189

\bibitem[{{Dodelson}(2003)}]{Dodelson:2003}
{Dodelson}, S. 2003, {Modern cosmology} (Modern cosmology / Scott
  Dodelson.~Amsterdam (Netherlands): Academic Press.~ISBN 0-12-219141-2, 2003,
  XIII + 440 p.)

\bibitem[{Donoho \& Huo(2001)}]{cur:donoho_01b}
Donoho, D. \& Huo, X. 2001, \itit, 47, 2845--2862

\bibitem[{{Doroshkevich} {et~al.}(2004){Doroshkevich}, {Tucker}, {Allam}, \&
  {Way}}]{SDSS:lrg2004}
{Doroshkevich}, A., {Tucker}, D.~L., {Allam}, S., \& {Way}, M.~J. 2004, \aap,
  418, 7

\bibitem[{{Douspis} {et~al.}(2008){Douspis}, {Castro}, {Caprini}, \&
  {Aghanim}}]{Douspis:2008}
{Douspis}, M., {Castro}, P.~G., {Caprini}, C., \& {Aghanim}, N. 2008, AAP, 485,
  395

\bibitem[{Efron(1979)}]{Efron1979}
Efron, B. 1979, Annals of Statistics, 7, 1

\bibitem[{Efron(1987)}]{Efron1987}
Efron, B. 1987, {The Jackknife, the Bootstrap, and Other Resampling Plans},
  CBMS-NSF Regional Conference Series in Applied Mathematics (Society for
  Industrial Mathematics)

\bibitem[{{Efstathiou}(2004)}]{Efstathiou2004hybrid}
{Efstathiou}, G. 2004, \mnras, 349, 603

\bibitem[{{Elad} {et~al.}(2005){Elad}, {Querre}, \& {Donoho}}]{Elad2005}
{Elad}, M.and~{Starck}, J.-L., {Querre}, P., \& {Donoho}, D. 2005, Applied and
  Computational Harmonic Analysis, 19, 340

\bibitem[{Elad {et~al.}(2005)Elad, Starck, Donoho, \&
  Querre}]{inpainting:elad05}
Elad, M., Starck, J.-L., Donoho, D., \& Querre, P. 2005, \acha, 19, 340--358

\bibitem[{Fadili {et~al.}(2009)Fadili, Starck, \& Murtagh}]{starck:jalal06}
Fadili, M.~J., Starck, J.-L., \& Murtagh, F. 2009, The Computer Journal, 52,
  64--79

\bibitem[{Flachaire(2005)}]{Flachaire2005}
Flachaire, E. 2005, Computational Statistics \& Data Analysis, 49, 361 , 2nd
  CSDA Special Issue on Computational Econometrics

\bibitem[{Fosalba \& Gazta\~naga(2004)}]{Fosalba:2004ge}
Fosalba, P. \& Gazta\~naga, E. 2004, \mnras, 350, L37

\bibitem[{Fosalba {et~al.}(2003)Fosalba, Gazta\~naga, \&
  Castander}]{Fosalba:2003ge}
Fosalba, P., Gazta\~naga, E., \& Castander, F. 2003, \apj, 597, L89

\bibitem[{{Francis} \&
  {Peacock}(2010{\natexlab{a}})}]{Francis:2010iswanomalies}
{Francis}, C.~L. \& {Peacock}, J.~A. 2010{\natexlab{a}}, \mnras, 406, 14

\bibitem[{{Francis} \&
  {Peacock}(2010{\natexlab{b}})}]{Francis:2010iswdetection}
{Francis}, C.~L. \& {Peacock}, J.~A. 2010{\natexlab{b}}, \mnras, 406, 2

\bibitem[{{Frommert} {et~al.}(2008){Frommert}, {En{\ss}lin}, \&
  {Kitaura}}]{Frommert:2008}
{Frommert}, M., {En{\ss}lin}, T.~A., \& {Kitaura}, F.~S. 2008, \mnras, 391,
  1315

\bibitem[{Gazta\~naga {et~al.}(2006)Gazta\~naga, Manera, \&
  Multamaki}]{Gaztanaga:2004sk}
Gazta\~naga, E., Manera, M., \& Multamaki, T. 2006, \mnras, 365, 171

\bibitem[{{Giannantonio} {et~al.}(2006){Giannantonio}, {Crittenden}, {Nichol},
  {Scranton}, {Richards}, {Myers}, {Brunner}, {Gray}, {Connolly}, \&
  {Schneider}}]{Giannantonio:2006al}
{Giannantonio}, T., {Crittenden}, R.~G., {Nichol}, R.~C., {et~al.} 2006, \prd,
  74, 063520

\bibitem[{{Giannantonio} {et~al.}(2008){Giannantonio}, {Scranton},
  {Crittenden}, {Nichol}, {Boughn}, {Myers}, \& {Richards}}]{Giannantonio:2008}
{Giannantonio}, T., {Scranton}, R., {Crittenden}, R.~G., {et~al.} 2008, \prd,
  77, 123520

\bibitem[{{G{\'o}rski} {et~al.}(2002){G{\'o}rski}, {Banday}, {Hivon}, \&
  {Wandelt}}]{healpix:2002}
{G{\'o}rski}, K.~M., {Banday}, A.~J., {Hivon}, E., \& {Wandelt}, B.~D. 2002, in
  Astronomical Society of the Pacific Conference Series, Vol. 281, Astronomical
  Data Analysis Software and Systems XI, ed. {D.~A.~Bohlender, D.~Durand, \&
  T.~H.~Handley}, 107--+

\bibitem[{Gorski {et~al.}(2005)}]{Gorski:2004by}
Gorski, K.~M. {et~al.} 2005, \apj, 622, 759

\bibitem[{{Granett} {et~al.}(2009){Granett}, {Neyrinck}, \&
  {Szapudi}}]{Granett:2009}
{Granett}, B.~R., {Neyrinck}, M.~C., \& {Szapudi}, I. 2009, \apj, 701, 414

\bibitem[{Hadamard(1902)}]{Hadamard1902}
Hadamard, J. 1902, {Sur les probl\`emes aux d\'eriv\'ees partielles et leur
  signification physique.}

\bibitem[{Hall(1995)}]{Hall1995}
Hall, P. 1995, {The Bootstrap and Edgeworth Expansion}, {Springer Series in
  Statistics} (Springer)

\bibitem[{{Hamimeche} \& {Lewis}(2008)}]{Hamimeche:2008}
{Hamimeche}, S. \& {Lewis}, A. 2008, \prd, 77, 103013

\bibitem[{{Hern{\'a}ndez-Monteagudo}(2008)}]{Hernandez:2008}
{Hern{\'a}ndez-Monteagudo}, C. 2008, \aap, 490, 15

\bibitem[{{Hern{\'a}ndez-Monteagudo}(2009)}]{Hernandez:2009}
{Hern{\'a}ndez-Monteagudo}, C. 2009, ArXiv e-prints

\bibitem[{{Hivon} {et~al.}(2002){Hivon}, {G{\'o}rski}, {Netterfield}, {Crill},
  {Prunet}, \& {Hansen}}]{Hivon:2002}
{Hivon}, E., {G{\'o}rski}, K.~M., {Netterfield}, C.~B., {et~al.} 2002, \apj,
  567, 2

\bibitem[{{Ho} {et~al.}(2008){Ho}, {Hirata}, {Padmanabhan}, {Seljak}, \&
  {Bahcall}}]{Ho:2008}
{Ho}, S., {Hirata}, C., {Padmanabhan}, N., {Seljak}, U., \& {Bahcall}, N. 2008,
  \prd, 78, 043519

\bibitem[{{Jarosik} {et~al.}(2010){Jarosik}, {Bennett}, {Dunkley}, {Gold},
  {Greason}, {Halpern}, {Hill}, {Hinshaw}, {Kogut}, {Komatsu}, {Larson},
  {Limon}, {Meyer}, {Nolta}, {Odegard}, {Page}, {Smith}, {Spergel}, {Tucker},
  {Weiland}, {Wollack}, \& {Wright}}]{Jarosik2010}
{Jarosik}, N., {Bennett}, C.~L., {Dunkley}, J., {et~al.} 2010, ArXiv e-prints

\bibitem[{Jarrett {et~al.}(2000)}]{Jarrett:2000me}
Jarrett, T.~H. {et~al.} 2000, Astron. J., 119, 2498

\bibitem[{{Kamionkowski} \& {Spergel}(1994)}]{Kamionkowski:1994s}
{Kamionkowski}, M. \& {Spergel}, D.~N. 1994, \apj, 432, 7

\bibitem[{{Kayo} {et~al.}(2001){Kayo}, {Taruya}, \& {Suto}}]{Suto:2001}
{Kayo}, I., {Taruya}, A., \& {Suto}, Y. 2001, \apj, 561, 22

\bibitem[{{Kinkhabwala} \& {Kamionkowski}(1999)}]{Kinkhabwala:1999k}
{Kinkhabwala}, A. \& {Kamionkowski}, M. 1999, Physical Review Letters, 82, 4172

\bibitem[{{Komatsu} {et~al.}(2009){Komatsu}, {Dunkley}, {Nolta}, {Bennett},
  {Gold}, {Hinshaw}, {Jarosik}, {Larson}, {Limon}, {Page}, {Spergel},
  {Halpern}, {Hill}, {Kogut}, {Meyer}, {Tucker}, {Weiland}, {Wollack}, \&
  {Wright}}]{Komatsu:2009}
{Komatsu}, E., {Dunkley}, J., {Nolta}, M.~R., {et~al.} 2009, \apjs, 180, 330

\bibitem[{Liu(1988)}]{Liu1988}
Liu, R.~Y. 1988, Annals of Statistics, 16, 1696

\bibitem[{{L{\'o}pez-Corredoira} {et~al.}(2010){L{\'o}pez-Corredoira}, {Sylos
  Labini}, \& {Betancort-Rijo}}]{Lopez:2010nocross}
{L{\'o}pez-Corredoira}, M., {Sylos Labini}, F., \& {Betancort-Rijo}, J. 2010,
  \aap, 513, A3+

\bibitem[{{Maddox} {et~al.}(1990){Maddox}, {Efstathiou}, {Sutherland}, \&
  {Loveday}}]{APM:1990}
{Maddox}, S.~J., {Efstathiou}, G., {Sutherland}, W.~J., \& {Loveday}, J. 1990,
  \mnras, 242, 43P

\bibitem[{Mammen(1993)}]{Mammen1993}
Mammen, E. 1993, Annals of Statistics, 21, 255

\bibitem[{{McEwen} {et~al.}(2007){McEwen}, {Vielva}, {Hobson},
  {Mart{\'{\i}}nez-Gonz{\'a}lez}, \& {Lasenby}}]{McEwen:2006md}
{McEwen}, J.~D., {Vielva}, P., {Hobson}, M.~P., {Mart{\'{\i}}nez-Gonz{\'a}lez},
  E., \& {Lasenby}, A.~N. 2007, \mnras, 376, 1211

\bibitem[{{McEwen} {et~al.}(2008){McEwen}, {Wiaux}, {Hobson}, {Vandergheynst},
  \& {Lasenby}}]{McEwen:2008}
{McEwen}, J.~D., {Wiaux}, Y., {Hobson}, M.~P., {Vandergheynst}, P., \&
  {Lasenby}, A.~N. 2008, \mnras, 384, 1289

\bibitem[{{Munshi} {et~al.}(2009){Munshi}, {Valageas}, {Cooray}, \&
  {Heavens}}]{Munshi2009gauss}
{Munshi}, D., {Valageas}, P., {Cooray}, A., \& {Heavens}, A. 2009, ArXiv
  e-prints

\bibitem[{Nolta {et~al.}(2004)}]{Nolta:2003uy}
Nolta, M.~R. {et~al.} 2004, \apj, 608, 10

\bibitem[{Padmanabhan {et~al.}(2005)}]{Padmanabhan:2004fy}
Padmanabhan, N. {et~al.} 2005, Phys. Rev., D72, 043525

\bibitem[{{Peacock} {et~al.}(2006){Peacock}, {Schneider}, {Efstathiou},
  {Ellis}, {Leibundgut}, {Lilly}, \& {Mellier}}]{WGFC}
{Peacock}, J.~A., {Schneider}, P., {Efstathiou}, G., {et~al.} 2006, {ESA-ESO
  Working Group on ''Fundamental Cosmology''}, Tech. rep.

\bibitem[{{Percival} {et~al.}(2007 a){Percival}, {Nichol}, {Eisenstein},
  {Weinberg}, {Fukugita}, {Pope}, {Schneider}, {Szalay}, {Vogeley}, {Zehavi},
  {Bahcall}, {Brinkmann}, {Connolly}, {Loveday}, \& {Meiksin}}]{Percival:2007b}
{Percival}, W.~J., {Nichol}, R.~C., {Eisenstein}, D.~J., {et~al.} 2007 a, \apj,
  657, 51

\bibitem[{{Perotto} {et~al.}(2010){Perotto}, {Bobin}, {Plaszczynski}, {Starck},
  \& {Lavabre}}]{perotto10}
{Perotto}, L., {Bobin}, J., {Plaszczynski}, S., {Starck}, J., \& {Lavabre}, A.
  2010, \aap, 519, A4+

\bibitem[{{Pietrobon} {et~al.}(2006){Pietrobon}, {Balbi}, \&
  {Marinucci}}]{Pietrobon:2006}
{Pietrobon}, D., {Balbi}, A., \& {Marinucci}, D. 2006, \prd, 74, 043524

\bibitem[{{Pires} {et~al.}(2009){Pires}, {Starck}, {Amara}, {Teyssier},
  {R{\'e}fr{\'e}gier}, \& {Fadili}}]{starck:pires08}
{Pires}, S., {Starck}, J., {Amara}, A., {et~al.} 2009, \mnras, 395, 1265

\bibitem[{{Pires} {et~al.}(2010){Pires}, {Starck}, \& {Refregier}}]{pires10}
{Pires}, S., {Starck}, J., \& {Refregier}, A. 2010, IEEE Signal Processing
  Magazine, 27, 76

\bibitem[{{Raccanelli} {et~al.}(2008){Raccanelli}, {Bonaldi}, {Negrello},
  {Matarrese}, {Tormen}, \& {de Zotti}}]{Raccanelli2008}
{Raccanelli}, A., {Bonaldi}, A., {Negrello}, M., {et~al.} 2008, \mnras, 386,
  2161

\bibitem[{Racine \& MacKinnon(2007)}]{Racine2007}
Racine, J.~S. \& MacKinnon, J.~G. 2007, Computational Statistics \& Data
  Analysis, 51, 5949

\bibitem[{{Rassat} {et~al.}(2007){Rassat}, {Land}, {Lahav}, \&
  {Abdalla}}]{Rassat:2007KRL}
{Rassat}, A., {Land}, K., {Lahav}, O., \& {Abdalla}, F.~B. 2007, \mnras, 377,
  1085

\bibitem[{{Rauhut} \& {Ward}(2010)}]{rauhut10}
{Rauhut}, H. \& {Ward}, R. 2010, ArXiv e-prints

\bibitem[{{Refregier} {et~al.}(2011){Refregier}, {Amara}, {Kitching}, \&
  {Rassat}}]{Refregier:2011}
{Refregier}, A., {Amara}, A., {Kitching}, T.~D., \& {Rassat}, A. 2011, \aap,
  528, A33+

\bibitem[{{Refregier} {et~al.}(2010){Refregier}, {Amara}, {Kitching}, {Rassat},
  {Scaramella}, {Weller}, \& {Euclid Imaging Consortium}}]{Euclidsb}
{Refregier}, A., {Amara}, A., {Kitching}, T.~D., {et~al.} 2010, ArXiv e-prints

\bibitem[{Sachs \& Wolfe(1967)}]{Sachs:1967er}
Sachs, R.~K. \& Wolfe, A.~M. 1967, \apj, 147, 73

\bibitem[{{Sawangwit} {et~al.}(2010){Sawangwit}, {Shanks}, {Cannon}, {Croom},
  {Ross}, \& {Wake}}]{Shanks:isw}
{Sawangwit}, U., {Shanks}, T., {Cannon}, R.~D., {et~al.} 2010, \mnras, 402,
  2228

\bibitem[{{Schaefer} {et~al.}(2010){Schaefer}, {Fotios Kalovidouris}, \&
  {Heisenberg}}]{Schaefer:2010}
{Schaefer}, B.~M., {Fotios Kalovidouris}, A., \& {Heisenberg}, L. 2010, ArXiv
  e-prints

\bibitem[{Schlegel {et~al.}(1998)Schlegel, Finkbeiner, \&
  Davis}]{Schlegel:1997yv}
Schlegel, D.~J., Finkbeiner, D.~P., \& Davis, M. 1998, \apj, 500, 525

\bibitem[{{Schrabback} {et~al.}(2010){Schrabback}, {Hartlap}, {Joachimi},
  {Kilbinger}, {Simon}, {Benabed}, {Brada{\v c}}, {Eifler}, {Erben},
  {Fassnacht}, {High}, {Hilbert}, {Hildebrandt}, {Hoekstra}, {Kuijken},
  {Marshall}, {Mellier}, {Morganson}, {Schneider}, {Semboloni}, {van Waerbeke},
  \& {Velander}}]{Schrabback:2010}
{Schrabback}, T., {Hartlap}, J., {Joachimi}, B., {et~al.} 2010, \aap, 516, A63+

\bibitem[{{Scranton} {et~al.}(2003)}]{Scranton:2003in}
{Scranton}, R. {et~al.} 2003, ArXiv Astrophysics e-prints

\bibitem[{{Song} {et~al.}(2007){Song}, {Hu}, \& {Sawicki}}]{Song:2007hs}
{Song}, Y.-S., {Hu}, W., \& {Sawicki}, I. 2007, prd, 75, 044004

\bibitem[{Spergel {et~al.}(2003)}]{Spergel:2003cb}
Spergel, D.~N. {et~al.} 2003, \apjs, 148, 175

\bibitem[{Starck {et~al.}(2010)Starck, Murtagh, \& Fadili}]{starck:book10}
Starck, J.-L., Murtagh, F., \& Fadili, M. 2010, Sparse Image and Signal
  Processing (Cambridge University Press)

\bibitem[{{Vielva} {et~al.}(2006){Vielva}, {Mart{\'{\i}}nez-Gonz{\'a}lez}, \&
  {Tucci}}]{Vielva2006}
{Vielva}, P., {Mart{\'{\i}}nez-Gonz{\'a}lez}, E., \& {Tucci}, M. 2006, \mnras,
  365, 891

\bibitem[{{Wild} {et~al.}(2005){Wild}, {Peacock}, {Lahav}, {Conway}, {Maddox},
  {Baldry}, {Baugh}, {Bland-Hawthorn}, {Bridges}, {Cannon}, {Cole}, {Colless},
  {Collins}, {Couch}, {Dalton}, {De Propris}, {Driver}, {Efstathiou}, {Ellis},
  {Frenk}, {Glazebrook}, {Jackson}, {Lewis}, {Lumsden}, {Madgwick}, {Norberg},
  {Peterson}, {Sutherland}, \& {Taylor}}]{Wild:2005}
{Wild}, V., {Peacock}, J.~A., {Lahav}, O., {et~al.} 2005, \mnras, 356, 247

\bibitem[{{Xia} {et~al.}(2009){Xia}, {Viel}, {Baccigalupi}, \&
  {Matarrese}}]{Xia2009}
{Xia}, J., {Viel}, M., {Baccigalupi}, C., \& {Matarrese}, S. 2009, \jcap, 9, 3

\end{thebibliography}

\end{document}